\documentclass[aps,prd,twocolumn,groupedaddress]{revtex4}

\pdfoutput=1
\usepackage {natbib}
\usepackage {graphicx}
\usepackage {color}

\usepackage{amsmath, amsthm, amsfonts}
\usepackage{hyperref}

\def\mbf#1{\ensuremath{\mathchoice{\mbox{\boldmath$\displaystyle#1$}}
{\mbox{\boldmath$\textstyle#1$}}
{\mbox{\boldmath$\scriptstyle#1$}}
{\mbox{\boldmath$\scriptscriptstyle#1$}}}}

\newcommand{\ms}{\mathrm{ms}}
\newcommand{\secs}{\mathrm{sec}}
\newcommand{\hours}{\mathrm{hours}}
\newcommand{\days}{\mathrm{days}}
\newcommand{\weeks}{\mathrm{weeks}}
\newcommand{\months}{\mathrm{months}}
\newcommand{\years}{\mathrm{yr}}
\newcommand{\kgmsqu}{\mathrm{kg}\,\mathrm{m}^2}

\newcommand{\Hz}{\mathrm{Hz}}
\newcommand{\erg}{\mathrm{erg}}

\newcommand{\Ord}[1]{\mathcal{O}\left(#1\right)}
\newcommand{\expect}[1]{E\left[#1\right]}

\newcommand{\prob}[2]{P\left(#1|#2\right)}
\newcommand{\scalar}[2]{\left(#1|#2\right)}

\newcommand{\detVec}[1]{\mbf{#1}}

\newcommand{\avg}[1]{\langle #1 \rangle}

\newcommand{\Uni}[2]{U\left(#1,\,#2\right)}

\newcommand{\normal}{{\mathrm{c}}}
\newcommand{\super}{{\mathrm{s}}}
\newcommand{\tGap}{{\Delta t}}
\newcommand{\glitch}{{\mathrm{glitch}}}
\newcommand{\quake}{{\mathrm{quake}}}
\newcommand{\tRelax}{\tau_\super}
\newcommand{\GW}{{\mathrm{GW}}}
\newcommand{\Om}{\Omega}

\newcommand{\hrmsSq}{\hrms^2}
\newcommand{\hrms}{\widehat{h}_0}
\newcommand{\hsd}{h_{\mathrm{sd}}}
\newcommand{\hrss}{h_{\mathrm{rss}}}

\newcommand{\sig}{\mathrm{s}}
\newcommand{\Dop}{\lambda}
\newcommand{\Freq}{f}

\newcommand{\nhat}{\hat{n}}

\newcommand{\Amp}{\mathcal{A}}
\newcommand{\cosi}{\cos\iota}
\newcommand{\phio}{\phi_0}

\newcommand{\hmax}{h_{\mathrm{max}}}
\newcommand{\rhoh}{\widehat{\rho}}
\newcommand{\rhohMax}{\rhoh_{\mathrm{max}}}

\newcommand{\parms}{\theta}	
\newcommand{\parmSpace}{\mathbb{P}}

\newcommand{\M}{\mathcal{M}}
\newcommand{\Minv}{\M}

\newcommand{\Sn}{S}	

\newcommand{\mis}{m}	
\newcommand{\Vol}{\mathcal{V}}
\newcommand{\Surf}{\mathbf{S}}

\newcommand{\dVx}{\detVec{x}}
\newcommand{\dVn}{\detVec{n}}
\newcommand{\dVh}{\detVec{h}}
\newcommand{\dVs}{\detVec{s}}

\newcommand{\rhoOpt}{\rho_o}

\newcommand{\Trans}{\mathcal{T}}
\newcommand{\win}{g}
\newcommand{\type}{\varpi}
\newcommand{\rect}{\mathrm{r}}	
\newcommand{\ex}{\mathrm{e}}	
\newcommand{\tStart}{t^{0}}
\newcommand{\tTau}{\tau}
\newcommand{\tTauDelta}{\tTau_\Delta}
\newcommand{\tTauMin}{\tTau_{\mathrm{min}}}

\newcommand{\efold}{3}

\newcommand{\Hyp}{\mathcal{H}}
\newcommand{\Noise}{{\mathrm{G}}}	
\newcommand{\Signal}{{\mathrm{S}}}
\newcommand{\HypS}{\Hyp_\Signal}
\newcommand{\HypN}{\Hyp_\Noise}

\newcommand{\OSN}{O_{\Signal\Noise}}	
\newcommand{\BSN}{B_{\Signal\Noise}}	
\newcommand{\BF}{B_{\F}}		
\newcommand{\Lr}{\mathcal{L}}		

\newcommand{\F}{\mathcal{F}}		
\newcommand{\Fmax}{\F_{\mathrm{max}}}
\newcommand{\ML}{\mathrm{ML}}	
\newcommand{\MP}{\mathrm{MP}}	

\newcommand{\probDet}{p_{\mathrm{det}}}	
\newcommand{\fA}{{\mbox{\tiny{FA}}}}	
\newcommand{\probfA}{p_{\fA}}		

\newcommand{\Inf}{\mathcal{I}}		

\newcommand{\Gauss}{\mathrm{Gauss}}

\newcommand{\BSNhmax}{\left.\BSN\right|_{\hmax}}
\newcommand{\BSNrhohMax}{\left.\BSN\right|_{\rhohMax}}

\newcommand{\rms}[1]{\mathrm{rms}\left[#1\right]}

\newcommand{\Tobs}{T_{\mathrm{obs}}}
\newcommand{\Tdata}{T_{\mathrm{data}}}

\newcommand{\rangeI}{\textit{range~I}}
\newcommand{\rangeII}{\textit{range~II}}
\newcommand{\rangeIII}{\textit{range~III}}
\newcommand{\tRef}{\tStart_{\mathrm{ref}}}

\newcommand{\Nseg}{N}
\newcommand{\Tseg}{\Delta T}
\newcommand{\iSeg}{{(i)}}
\newcommand{\scoh}{\mathrm{sc}}
\newcommand{\SF}{\mathcal{S}_{\!\Nseg}}

\newcommand{\NSFT}{{N_{\mathrm{SFT}}}}
\newcommand{\TSFT}{{T_{\mathrm{SFT}}}}
\newcommand{\NStart}{{N_{\tStart}}}
\newcommand{\NTau}{{N_{\tTau}}}
\newcommand{\CFS}{{\texttt{CFSv2}}}

\newcommand{\atoms}{\mathrm{atoms}}
\newcommand{\Fmap}{{\F\mathrm{map}}}

\newcommand{\cost}{\mathfrak{c}}
\newcommand{\marg}{\mathrm{marg}}

\newcommand{\dcc}{LIGO-P1100002-v2}

\def\commitID{commitID: a2869dfcb32809ba959c095cee8213749caf261a}
\def\commitDATE{ Sat Apr 9 14:56:09 2011 +0200}

\def\commitSTATUS{CLEAN}

\hypersetup{
  pdftitle={Search method for long-duration gravitational-wave transients from neutron stars},
  pdfauthor={R Prix, S Giampanis, C Messenger},
  pdfkeywords={\$\commitID-\commitSTATUS{} \$}
}

\begin{document}

\title{Search method for long-duration gravitational-wave transients from neutron stars}

\author{R.~Prix}
\email{Reinhard.Prix@aei.mpg.de}
\author{S.~Giampanis}
\email{gstef@gravity.phys.uwm.edu}
\author{C.~Messenger}
\email{chris.messenger@astro.cf.ac.uk}
\affiliation{Albert-Einstein-Institut, Callinstr 38, 30167 Hannover, Germany}
\date{\commitDATE\\\mbox{\dcc}}

\begin{abstract}
We introduce a search method for a new class of gravitational-wave signals, namely long-duration
$\Ord{\hours - \weeks}$ transients from spinning neutron stars.
We discuss the astrophysical motivation from glitch relaxation models and we derive a
rough estimate for the maximal expected signal strength based on the superfluid excess rotational energy.
The transient signal model considered here extends the traditional class of infinite-duration
\emph{continuous-wave} signals by a finite start-time and duration.
We derive a multi-detector Bayes factor for these signals in Gaussian noise using $\F$-statistic amplitude
priors, which simplifies the detection statistic and allows for an efficient implementation.
We consider both a fully coherent statistic, which is computationally limited to directed searches for
known pulsars, and a cheaper semi-coherent variant, suitable for wide parameter-space searches for transients
from unknown neutron stars.
We have tested our method by Monte-Carlo simulation, and we find that it outperforms orthodox
maximum-likelihood approaches both in sensitivity and in parameter-estimation quality.
\end{abstract}

\maketitle

\section{Introduction}\label{sec:introduction}

Gravitational waves (GWs), predicted by Einstein's General Theory of Relativity, are expected to be emitted
by spinning neutron stars (NSs) with non-axisymmetric deformations or currents. Traditionally we distinguish
two categories of GW emission from NSs:

(i) short ``burst-like'' GWs from NS oscillations, e.g.\ fundamental (f-), pressure
(p-), Rossby (r-) or w-modes, which could be excited by a NS glitch. These signals would
typically be in the kHz frequency range and would be damped on timescales of milliseconds
(see \cite{andersson98:_towards_gw_asteroseism,2001MNRAS.320..307K}).
Various data-analysis methods for such high-frequency bursts have been developed (e.g.\ see
\cite{2007PhRvD..76d3003C,2008CQGra..25r4016H}) and recently a search for a GW burst from a glitch
in the Vela pulsar has been performed on LIGO data \cite{clarkLSC2010:_vela_glitch}.

(ii) long-duration ``continuous waves''(CWs) from non-axisymmetric deformations or currents in spinning NSs.
These CWs are quasi-sinusoidal with a well-defined, slowly varying frequency $\Freq$,
which is typically of order of the NS spin frequency $\nu$: in particular $f=2\,\nu$ for ``mountains'' or
precession, additionally $f\approx \nu$ for precessing NSs, while $f\approx 4\nu/3$ for r-mode oscillations.
In the past decade a number of data-analysis methods have been developed and applied to searches for CW
signals in the data of ground-based detectors. These searches typically come in two flavors, either fully coherently
targeting known pulsars at $\Freq=2\nu$, or semi-coherently searching for CWs from unknown NSs in a large parameter
space of frequencies and sky-positions. See \cite{prix06:_cw_review} for a review of the astrophysical models,
data-analysis methods and results of searches for CWs, and for further references.

The traditional CW model assumes that these signals are truly \emph{continuous} in the sense of a
quasi-infinite duration, or at least of longer duration than the available observation time, typically $\Tobs\sim1-2\,\years$.
This postulates a spinning neutron star with a quasi-stable non-axisymmetric deformation, such that the
relaxation to the axisymmetric thermodynamic equilibrium happens on long timescales $\gg \Tobs$.
Alternatively, the non-axisymmetry can be driven by external influences such as accretion in low-mass
X-ray binary systems (LMXBs), or by GW-driven instabilities such as unstable $r$-mode oscillations.
Both are complex dynamical processes that need to be perfectly stable in order to produce a traditional CW
signal.

In the wide gap between burst-like $\Ord{\ms}$ and truly continuous $\Ord{\infty}$ signals,
spinning NSs can reasonably be assumed to emit ``CW'' signals of intermediate duration of
$\Ord{\hours-\weeks}$.
We refer to this third category of GWs from spinning NSs by the oxymoron ``transient CWs''.
We can give three plausibility arguments for why it would be worthwhile to explore this new parameter space:
(a) Young NSs exhibit enigmatic sudden spin-up events called ``glitches'', followed by a relaxation phase
with timescale between days to months. This shows that internal dynamical processes on these timescales do
exist in NSs.
(b) Equilibrium NS configurations are axisymmetric and GWs from quasi-stable deviations are expected to be weak:
deformations strong enough to produce detectable GWs on Earth would therefore seem more likely to be
associated with ``catastrophic'' transient events.
(c) Our astrophysical understanding of the universe is incomplete. Independently of any astrophysical models,
this range of parameter space is currently not covered by any other searches and should therefore be investigated.

The transient-CW signal model is very similar to the standard CW model, namely quasi-sinusoidal emission with a
well-defined, slowly varying frequency of the order of the NS spin frequency, $\Freq \sim \Ord{\nu}$.
Contrary to CWs, however, the signal has a definite start-time $\tStart$ and a finite effective lifetime $\tRelax$, and
the signal amplitude $h_0$ can be modulated by a window-function, for example an exponential decay.

While the present study is concerned with extending the traditional $\Ord{\infty}$ CW search to finite durations,
there is an independent and complementary effort underway (called ``STAMP'') to extend the traditional $\Ord{\ms}$ ``burst''
searches to longer durations \cite{thrane2010:_intermediatePipeline}.
This ``long burst'' method does not assume a parametrized signal model, but instead tries to find connected
time-frequency patterns in the cross-correlation power between multiple detectors. These transients
can span any duration from seconds to weeks. This approach will be computationally cheaper and more robust
towards a wider class of unmodeled GW signals, but would therefore also be expected to be less
sensitive towards the particular class of transient-CW signals considered here.

The plan of this paper is as follows: in
Sec.~\ref{sec:astr-toy-model} we discuss the astrophysical motivation for transient CWs, including a simple
signal-to-noise ratio (SNR) estimate based on the superfluid excess rotation energy in NSs.
Sec.~\ref{sec:model} introduces the parametrized transient-CW signal model, and
Sec.~\ref{sec:detect-meth-bayes} develops the coherent and semi-coherent Bayesian search methods for these signals.
Sec.~\ref{sec:detection-efficiency} presents numerical Monte-Carlo results on the detection efficiency, and
Sec.~\ref{sec:parameter-estimation} illustrates the performance on parameter estimation.
Sec.~\ref{sec:Conclusions} gives a concluding summary.
Details on implementation and computing cost estimates are given in the appendix~\ref{sec:trans-search-impl}.

\section{Astrophysical motivation for transient CWs}
\label{sec:astr-toy-model}

\subsubsection*{Concrete astrophysical predications}

There are currently very few concrete predictions in the literature regarding the existence and
properties of potential transient-CW signals from spinning NSs.
Notable exceptions are the recent studies \cite{2008CQGra..25v5020V,2010MNRAS.409.1705B} of
GW emission from non-axisymmetric Ekman flow during the post-glitch relaxation phase, which typically lasts
for days to months (e.g.\ see \cite{2000MNRAS.315..534L}). The authors conclude that GWs from this mechanism
could be detectable with second- or third-generation ground-based detectors.
Another interesting recent idea \cite{2011arXiv1102.4830K} suggests that giant magnetic flares in magnetars
could trigger polar Alfv\'en oscillations, emitting GWs at around $\sim100$~Hz and lasting for days to months,
although the details of the underlying physics are uncertain \cite{2011arXiv1103.0880L}.

One can further speculate on a number of potential transient-CW emission mechanisms from spinning NSs.
Many ``classical'' CW mechanisms discussed in the literature (e.g.\ see \cite{prix06:_cw_review} for references)
typically have large uncertainties on the lifetime of the emission, and are therefore potential sources of
transient-CW signals.
For example, free precession of NSs occurs when the spin-axis is misaligned with the axis of symmetry, and has
long been considered a possible mechanism for CW emission (e.g.\ \cite{jks98:_data}). A more detailed
analysis \cite{jones02:_gravit} concluded that the emission from free precession could be damped on a
(highly uncertain) timescale of a few weeks to years.
Young NSs could be deformed by extreme magnetic fields and result in stronger transient CWs from damped free
precession \cite{2010arXiv1011.2778G}.
Newly-born magnetars with strong toroidal fields could be subject to a magnetic instability producing
strong GWs on the timescale of several days \cite{2009MNRAS.398.1869D}.
Similarly, the strength and timescale of the r-mode GW instability remain highly uncertain (e.g.\ see
\cite{2001IJMPD..10..381A} for a review), despite a number of studies over more than a decade. This
instability could operate on timescales of days to months in newly-born NS. As an example, see
\cite{2009PhRvD..79j4003B} for recent work on the large variety of possible scenarios for the r-mode
instability and spindown-evolution taking into account nonlinear mode couplings.

\subsubsection*{Energetics and SNR of transient CWs}

It is instructive to consider the general relation between GW energy emitted and the average expected SNR for
transient CWs. This can be useful to derive an order-of-magnitude estimate of the expected transient-CW SNR in a
simple toy-model.

In the following estimate we assume that the CWs are emitted by a non-axisymmetric deformation
$\epsilon(t)$ of the quadrupole moment, emitting GWs at frequency $\Freq$.
See also \cite{2010PhRvD..82j4002O} for a discussion of the relevant relations in the case of r-modes.
The total energy emitted in GWs is $E_\GW=\int L_\GW\,dt$, in terms of
the GW luminosity $L_\GW$ (e.g.\ see \cite{prix06:_cw_review}), which is
\begin{equation}
  \label{eq:83}
  L_\GW(t) = \frac{1}{10}\frac{G}{c^5}(2\pi\Freq)^6\,I^2\,\epsilon^2(t)\,,
\end{equation}
where $I$ is the axial moment of inertia, and $\epsilon(t)$ is the dimensionless deviation from axisymmetry of
the spinning NS. $G$ is Newton's gravitational constant, and $c$ is the speed of light.
We can write the corresponding signal amplitude $h_0(t)$ at the observer as
\begin{equation}
  \label{eq:89}
  h_0(t) = \frac{4\pi^2 G}{c^4} \frac{I\,\Freq^2}{d}\,\epsilon(t)\,,
\end{equation}
where $d$ is the distance to the NS.
Combining Eqs.~\eqref{eq:83} and \eqref{eq:89}, one can eliminate $\epsilon(t)$ and write the total GW energy
emitted during a time span $T$ as
\begin{equation}
  \label{eq:91}
  E_\GW = \frac{2\pi^2\,c^3}{5\,G}\,\Freq^2\,d^2\, \int^{T} h_0^2(t)\,dt\,,
\end{equation}
assuming a roughly constant average frequency $\Freq$.
The expected optimal signal-to-noise ratio (SNR), which will be defined in Eq.~\eqref{eq:35}, can be averaged
over sky-position and signal polarizations \cite{jks98:_data,prix:_cfsv2}, to yield
\begin{equation}
  \label{eq:90}
  \avg{\rhoOpt^2} = \frac{4}{25}\,\frac{1}{\Sn(\Freq)}\,\int^T h_0^2(t)\,dt\,,
\end{equation}
where $\Sn(\Freq)$ is the (single-sided) noise power spectral density (at the signal frequency $\Freq$) of the detector (or multi-detector combination,
see \cite{cutler05:_gen_fstat,prix:_cfsv2}).
By combining Eqs.~\eqref{eq:91} and \eqref{eq:90}, we obtain the average optimal SNR of a transient CW
in terms of the total emitted GW energy $E_\GW$, namely
\begin{equation}
  \label{eq:82}
  \avg{\rhoOpt^2} = \frac{2 G}{5\pi^2\,c^3}\,\frac{E_\GW}{\Sn(\Freq)\,\Freq^2\,d^2}\,,
\end{equation}
which agrees with the result given in \cite{2002grg..conf...72C}.
An analogous expression was derived in \cite{2002CQGra..19.1247O} for the case of r-modes.
This expression is interesting for two reasons:
(i) at fixed frequency $\Freq$, the expected SNR only depends on the total GW energy emitted, and not on the
timescale of the emission, i.e.\ short-strong or long-weak transient CWs of the same total energy typically result in
the same SNR, and
(ii) at fixed emitted $E_\GW$, the optimal SNR $\rhoOpt$ \emph{decreases} linearly with increasing GW
frequency $\Freq$, i.e.\ for the same transient GW energy, transients from slowly rotating NS would be easier
to detect than from fast rotators.

For convenience we define the dimensionless root-mean-square amplitude $\hrms$ over the timescale
$T$ as
\begin{equation}
  \label{eq:104}
  \hrmsSq\,T \equiv \int^{T} h_0^2(t)\,dt\,.
\end{equation}
Note that this quantity is not to be confused with the ``root-sum-squared'' amplitude
$\hrss$ often used to characterize signal strength in the context of burst searches (e.g.\ see
\cite{2007PhRvD..76d3003C}), which is defined as $\hrss^2\equiv\int h^2(t)\,dt$. The difference
is that $\hrss$ has dimension $\sqrt{t}$ and refers to the \emph{measured strain} $h(t)$ in a given detector (see
Eq.~\eqref{eq:1}), which is rapidly oscillating at frequency $\Freq$.
The dimensionless $\hrms$, on the other hand, refers to the intrinsic signal amplitude $h_0(t)$, which for
a transient CW would be slowly varying on a timescale $T$.
From the definition \eqref{eq:35} of the (optimal) SNR, one can see that $\rhoOpt^2=(2/\Sn)\hrss^2$, and
therefore Eq.~\eqref{eq:90} also yields the relation
\begin{equation}
  \label{eq:103}
  \avg{\hrss^2} = \frac{2}{25}\,\hrmsSq\,T\,.
\end{equation}
The quantity $\hrmsSq\,T$ is proportional to the GW energy \eqref{eq:91}, as well as the average optimal
SNR \eqref{eq:90}, namely
\begin{equation}
  \label{eq:96}
  \hrmsSq\,T \propto \avg{\rhoOpt^2}\,\Sn \propto \frac{E_\GW}{\Freq^2\,d^2}\,.
\end{equation}

\subsubsection*{Transient CWs from superfluid excess energy}

As a simple toy-model for the energy available \emph{in principle} for transient CWs from a spinning NS,
we consider the standard two-fluid model, which is at the core of current attempts to understand pulsar glitches.
See \cite{2000MNRAS.315..534L} for an overview of the phenomenology of observed pulsar glitches and the basic
two-fluid model.

The observed pulses (arriving with frequency $\nu$) are commonly associated with the rotating NS magnetic field, which is
attached to the crust and the \emph{normal} fluid interior, both rotating with angular velocity $\Om = 2\pi\nu$.
This normal component has a moment of inertia $I_\normal$, believed to form the bulk of the total moment of
inertia, $I \approx I_\normal$.
The basic two-fluid model assumes that some of the interior neutrons are superfluid, forming an independent
component that rotates at an (unobserved) angular velocity $\Om_\super$, and which has a moment of inertia $I_\super$, typically believed to
be of order $I_\super/I_\normal \sim 10^{-2}$ (e.g.\ see \cite{2000MNRAS.315..534L}).
The normal NS components slow down at an observed rate $\dot{\Om} = 2\pi\dot{\nu}$ due to
losses of angular momentum from the electromagnetic emission or interactions with the surrounding medium.
The superfluid, on the other hand, is believed to be weakly coupled to the normal component (therefore
$\dot{\Om}_\super\approx0$) and would continue to spin with angular velocity $\Om_\super$, until the ``lag''
$\Delta\Om = \Om_\super - \Om = - \dot{\Om}\,\tGap$ between the two components reaches a critical level (see
\cite{2010MNRAS.405.1061S} for a more detailed study including superfluid coupling effects). The timescale
$\tGap$ here could correspond to the inter-glitch period, if one assumes that every glitch restores perfect
co-rotation between the fluids.
At this ``breaking point'' some type of instability is believed to occur, resulting in the transfer of
angular momentum from the superfluid to the normal fluid. This would produce the visible ``glitch'', i.e.\ a
sudden spin-up $\delta\Om = 2\pi\delta\nu$ of the observed pulse frequency $\nu$.
The details of this instability are poorly understood and various models have been suggested in the
literature, such as the crust breaking due to the strain exerted by pinned vortices (e.g.\ see
\cite{1991ApJ...382..587R}), the vortex array becoming unpinned due to the Magnus force (e.g.\ see
\cite{1991ApJ...373..592L}), or a two-stream instability developing due to the dynamical coupling of the two
fluids \cite{acp03:_twostream_prl,acp04:_twostream}. See also \cite{carter00:_centr_buoyancy} for
an alternative mechanism to vortex pinning that can produce crust strain.
Similarly unclear are the physical mechanisms involved in the ensuing ``relaxation'' back to a state of
pre-glitch steady-state spin-down, which typically occurs on timescales of days to months (see
\cite{2000MNRAS.315..534L}).

During a glitch, the observed normal fluid would spin up by $\delta\Om$, while the superfluid would spin
down by $\delta\Om_\super$, such that angular momentum is conserved. Following \cite{2010MNRAS.405.1061S} and
assuming that the moments of inertia are constant during a glitch, we have
\begin{equation}
  \label{eq:45}
  I_\normal\, \delta\Om + I_\super\,\delta\Om_\super = 0\,,
\end{equation}
where typically the observed spin-up is up to $\delta\Om/\Om\sim 10^{-6}$. Note that for fiducial values of
$I_\super/I_\normal\sim10^{-2}$ this implies that the superfluid spins down by $\delta\Om_\super/\Om\sim -10^{-4}$ during a glitch.
Besides this angular-momentum transfer in the glitch, there is some excess energy $E_\glitch$ left.
Assuming that co-rotation between the two fluids is restored after the glitch, namely $\delta\Om_\super = -\Delta\Om + \delta\Om$,
the glitch excess energy can be shown to be
\begin{equation}
  \label{eq:67}
  E_\glitch = \frac{1}{2}\,I_\super\,\delta\Om_\super^2 + \frac{1}{2}\,I_\normal\,\delta\Om^2
  \approx \frac{1}{2}\,I_\super\,\delta\Om_\super^2\,,
\end{equation}
which agrees with the result in \cite{2010MNRAS.405.1061S}, where in the last step we assumed
the fiducial scales $I_\super/I_\normal\sim10^{-2}$ and therefore $\delta\Om/\delta\Om_\super\sim10^{-2}$,
which makes the second term negligible.
Using Eq.~\eqref{eq:45}, this can also be expressed as $E_\glitch \approx
\frac{1}{2}I_\normal\,\delta\Om\,\Delta\Om$.
This energy would have to be dissipated directly in the glitch, for example by exciting oscillation modes, which
would radiate GWs on short timescales, as considered in \cite{2007PhRvD..76d3003C,clarkLSC2010:_vela_glitch},
or by heating up the NS (e.g.\ see \cite{1991ApJ...381L..47V}).

However, after a glitch the NS generally relaxes back to a steady-state spindown on a timescale of days to
months, and therefore some (if not all) of the initial excess energy stored in the faster-rotating superfluid
could be driving GW emission on this \emph{relaxation} timescale of $\Ord{\days-\months}$. The recent studies
of post-glitch spin-up of the fluid NS core by non-axisymmetric Ekman flow \cite{2008CQGra..25v5020V,2010MNRAS.409.1705B}
provide one concrete example for exactly such a mechanism.
It is also conceivable that this hidden energy in the superfluid ``flywheel'' can lead to transient
CW emission directly, via an internal instability, without transferring its angular momentum to the curst
first, i.e.\ without a visible pulsar glitch.

The well-known spindown upper limit on CW emission from known pulsars \cite{prix06:_cw_review} assumes that
the total steady-state spindown energy of the pulsar is emitted as GWs. Similarly we can consider a
``superfluid upper limit'' on transient CWs, where the total superfluid excess energy is converted into
transient CWs, for example by sustaining some non-axisymmetric process that leads to emission at
frequencies $\Freq\sim\Ord{\nu}$.
The excess rotational energy $E_\super$ of the superfluid is
\begin{equation}
  \label{eq:61}
  E_\super = \frac{1}{2}\,I_\super\,(\Om_\super^2 - \Om^2) \approx 4\pi^2\,I_\super\,\nu\,\Delta\nu\,,
\end{equation}
where in the last expression we dropped the second-order energy term \eqref{eq:67}, which is smaller by
$\Delta\Omega/\Omega \approx 10^{-4}$. In the following we return to using the spin frequency $\nu$ instead of
the angular velocity $\Om = 2\pi\,\nu$.
Note that this expression happens to be numerically very similar to the ``starquake''
glitch-energy derived under the assumption of a spin-up caused by a reduction in the crust's moment of
inertia (e.g.\ see \cite{clarkLSC2010:_vela_glitch}), which results in
$E_{\quake} = 2\pi^2\,I\,\nu\,\delta\nu$, and substituting $\delta\nu= (I_\super/I)\Delta\nu$ from \eqref{eq:45}, we
obtain $2 E_{\quake} = E_\super$. For fiducial values of $I\sim10^{38}\kgmsqu$, Vela spin frequency $\nu\sim 10\,\Hz$
and a large glitch of $\delta\nu/\nu \sim 10^{-6}$, this yields $E_{\super}\sim 4\times10^{42}\,\erg$.

We have no direct observational evidence about the size of the lag $\Delta\Om = 2\pi\Delta\nu$ or the superfluid
angular velocity $\Om_\super$. In the simplest models one typically assumes the lag to be reset to zero after
every glitch, starting to build up again due to spindown $\dot{\nu}$. However, this is not
necessarily the case and the lag could accumulate over longer timescales and be correspondingly larger.

To obtain an upper limit estimate on the signal strength, we can equate $E_\GW$ of Eq.~\eqref{eq:91} with
$E_\super$ given in Eq.~\eqref{eq:61}, assuming that the GW is emitted at a frequency $\Freq=2\nu$
(corresponding to a non-axisymmetric deformation). This corresponds to the extreme case where the built-up
excess superfluid rotational energy drives emission of a transient CW. Combining this with Eq.~\eqref{eq:104},
we can obtain a superfluid transient-CW upper limit estimate in the form
\begin{equation}
  \label{eq:92}
  \hrms\sqrt{T} = \frac{1}{d}\sqrt{ \frac{5 G}{2 c^3} \, I_\super\,\frac{\Delta\nu}{\nu}}\,.
\end{equation}
Alternatively, if we assume that the total excess angular momentum is transferred in a glitch, we can use
Eq.~\eqref{eq:45} to rewrite this in terms of more directly observed glitch quantities, namely
\begin{equation}
  \label{eq:93}
  \hrms\sqrt{T} = \frac{1}{d}\sqrt{ \frac{5 G}{2 c^3} \, I\,\frac{\delta\nu}{\nu}}\,,
\end{equation}
where $I$ is the axial NS moment of inertia, and $\delta\nu/\nu$ is the observed glitch spin-up.
As a third alternative, we can use the simple two-fluid model for the buildup of the lag,
namely $\Delta\nu = |\dot{\nu}|\tGap$, and parametrize the unknown superfluid moment of inertia via
$I_\super = \gamma\,I$ (with fiducial value $\gamma\sim10^{-2}$), and obtain the relation
\begin{equation}
  \label{eq:94}
  \hrms = \sqrt{ \frac{ \gamma\,\tGap }{T} }\; \hsd \,,
\end{equation}
in terms of the well-known ``spindown limit'' $\hsd$, given by
\begin{equation}
  \label{eq:95}
  \hsd = \frac{1}{d}\sqrt{ \frac{5 G}{2 c^3} \, I\,\frac{|\dot{\nu}|}{\nu}}\,,
\end{equation}
which is derived from the assumption that the NS spindown energy is completely converted into GWs.
The relation \eqref{eq:94} shows that the superfluid transient-CW upper limit can be similar or even higher
than the usual pulsar spindown limit: a fraction $\gamma$ of the steady-state spindown energy is
accumulated in the superfluid over an inter-glitch timescale $\tGap$, and is released on a short timescale
$T$. Assuming $\gamma\sim10^{-2}$, inter-glitch periods of $\tGap\sim 1\,\years$ and transient-CW timescale of
$T\sim 2\,\weeks $, we see from Eq.~\eqref{eq:94} that for these values $\hrms \approx \hsd$.
This is interesting as it indicates that the young (and glitching) pulsars with the most promising spindown
upper limits on CW emission might also be the most attractive targets for directed transient-CW searches, for
example the Crab pulsar, Vela, and J0537-69 (e.g.\ see Fig.~4 in \cite{prix06:_cw_review}).

\section{Signal model for transient CWs} \label{sec:model}

The family of transient-CW signals considered here is a straightforward generalization of the traditional
infinite-duration CW model (\cite{brady98:_search_ligo_periodic,jks98:_data}, allowing for a finite
duration and non-trivial time-evolution of the overall amplitude $h_0$.
The set $\Trans$ of extra transient-CW parameters therefore consists of the start-time $\tStart$, characteristic timescale
$\tTau$, and the type of window function $\type$, i.e.\ $\Trans \equiv \{\type,\tStart,\tTau \}$.
The transient-CW signal family therefore simply consists of a window-function $\win_\type(t;\tStart,\tTau)$ applied to the
standard CW signal model, namely
\begin{equation}
  \label{eq:1}
  h^X(t;\Amp,\Dop,\Trans) = \win_\type(t;\tStart,\tTau)\;\Amp^\mu\,h^X_\mu(t;\Dop)\,,
\end{equation}
were we use implicit summation over the four amplitudes, $\mu=1,\dots 4$, and $X$ is an index over different
detectors.
The four canonical amplitudes $\Amp^\mu$ are functions of the CW amplitude $h_0$, polarization angles
$\cosi,\psi$ and the initial phase $\phio$, i.e.\ $\Amp^\mu = \Amp^\mu(h_0,\cosi,\psi,\phio)$. The
corresponding basis functions $h^X_\mu(t;\Dop)$ are found, for example, in \cite{PrixWhelan07:_MLDC1}, but
their explicit form is not relevant to our discussion here. The set of Doppler parameters $\Dop$
determines the time evolution of the signal phase, for example the source sky-position $\nhat$ and the GW
frequency $\Freq(t)$, which is generally allowed to be slowly varying with time. If the CW source is a neutron
star in a binary system, $\Dop$ would also include the relevant orbital parameters of the system.

In the following we restrict our attention to two simple types of transient window functions $\win_\type(t)$, namely
either \emph{rectangular}, denoted as $\type=\rect$, i.e.\
\begin{equation}
  \label{eq:2}
  \win_\rect(t;\tStart,\tTau) \equiv \left\{\begin{array}{ll}
      1 & \mbox{if } t \in [\tStart,\tStart+\tTau]\\
      0 & \mbox{otherwise}\,,
      \end{array}\right.
\end{equation}
or \emph{exponentially decaying}, denoted as $\type=\ex$, namely
\begin{equation}
  \label{eq:3}
  \win_\ex(t;\tStart,\tTau) \equiv \left\{\begin{array}{ll}
      e^{-(t - \tStart)/\tTau} & \mbox{if } t \in [\tStart,\tStart+\efold\tTau]\\
      0 & \mbox{otherwise}\,,
      \end{array}\right.
\end{equation}
where we somewhat arbitrarily truncated the exponential window at an e-folding of $\efold$, in order to
simplify the practical implementation of this window. This truncation gives the window a finite duration of $\efold\tTau$,
and at the truncation-point the amplitude has decreased by more than $95\%$ and we can neglect the
corresponding loss of SNR.

Following the notation of \cite{cutler05:_gen_fstat,prix06:_searc}, we use boldface to
denote multi-detector vectors, i.e.\ $\{\dVx\}^X = x^X$ denotes the set of data-streams $x(t)$ from different
detectors $X$. We can now conveniently absorb the window-function $\win(t)$ in Eq.~\eqref{eq:1} into the definition of
new transient basis functions $\dVh_\mu'$, namely
\begin{equation}
  \label{eq:14}
  \dVh'_\mu(t; \Dop,\Trans) \equiv \win_\type(t;\tStart,\tTau)\,\dVh_\mu(t;\Dop)\,.
\end{equation}
If we denote $\parms$ the set of \emph{all} signal parameters of our search, i.e.\
\begin{equation}
  \label{eq:6}
  \parms \equiv \{\Amp, \Dop, \Trans \}\,,
\end{equation}
then we can write the transient signal model \eqref{eq:1} now more compactly as
\begin{equation}
  \label{eq:5}
  \dVh(t;\parms) = \Amp^\mu\,\dVh'_\mu(t;\Dop, \Trans)\,.
\end{equation}

\section{Detection method: Odds ratio}
\label{sec:detect-meth-bayes}

\subsection{Hypothesis testing framework}
\label{sec:hypoth-test-fram}

Based on observed data $\dVx$, we want to decide between two hypotheses: under the noise hypothesis
$\HypN$ the observed data consists only of Gaussian stationary noise $\dVn$, and under the signal
hypothesis $\HypS$ the data contains in addition a transient-CW signal $\dVh(t;\parms)$ of Eq.~\eqref{eq:5},
namely
\begin{align}
  \HypN:  \dVx(t) &= \dVn(t)\,,   \label{eq:7a}\\
  \HypS: \dVx(t) &= \dVn(t) + \dVh(t;\parms)\quad \mbox{for any}\quad\parms\in\parmSpace\,,\label{eq:7b}
\end{align}
where $\parmSpace$ denotes the signal parameter space.
Note that the signal hypothesis \eqref{eq:7b} is incomplete without the specification of a
probability distribution for the unknown signal parameters $\parms$ over their parameter space $\parmSpace$, i.e.\ a \emph{prior}
probability $\prob{\parms}{\HypS}$. Note that for simplicity our notation does not explicitly distinguish
between proper probabilities and probability densities. This difference is implicit in the type of argument,
namely whether it is discrete, such that $\sum_i \prob{x_i}{\Inf} = 1$, or continuous, where $\int \prob{x}{\Inf}\,d x = 1$.
Furthermore, we sometimes (but not always) explicitly state $\Inf$ as a conditional in probability statements, denoting
the full set of remaining relevant prior model assumptions that the probability statement depends on.

\subsection{Gaussian noise and scalar product}
\label{sec:gauss-noise-scal}

For quasi-sinusoidal CWs \eqref{eq:1} in stationary Gaussian noise we can define a multi-detector scalar product
\cite{cutler05:_gen_fstat}, using the notation of \cite{prix06:_searc}, as
\begin{equation} \label{eq:inner_product}
  \scalar{\dVx}{\detVec{y}} \equiv 2 \sum_X \Sn^{-1}_X(f) \int x^X(t) \, y^X(t) \,dt \,,
\end{equation}
where $\Sn_X(f)$ is the (stationary) single-sided noise power spectral density in detector $X$, which is
assumed constant over a narrow frequency band around the signal frequency $f$.
This allows us to write the likelihood for the data $\dVx$ in the Gaussian noise-case \eqref{eq:7a} as
\begin{equation}
  \label{eq:4}
  \prob{\dVx}{\HypN} = \kappa\,e^{-\frac{1}{2}\scalar{\dVx}{\dVx}}\,,
\end{equation}
where $\kappa$ is a normalization constant. In the presence of a signal $\dVh(t;\parms)$ with parameters $\parms$,
subtracting this signal from the data $\dVx$ results again in pure Gaussian noise $\dVn$,
i.e.\ $\dVn = \dVx - \dVh(\parms)$, and therefore
\begin{equation}
  \label{eq:7}
  \prob{\dVx}{\HypS,\parms} = \kappa \, e^{-\frac{1}{2}\scalar{\dVx - \dVh(\parms)}{\dVx - \dVh(\parms)}}\,.
\end{equation}
The likelihood for the data $\dVx$ containing \emph{any} signal $\dVh(t;\parms)$ with $\parms \in \parmSpace$
drawn from the prior $\prob{\parms}{\HypS}$ can easily be obtained (e.g.\ see \cite{2009CQGra..26t4013P}) as
\begin{equation}
  \label{eq:8}
  \prob{\dVx}{\HypS} = \int_\parmSpace \prob{\dVx}{\HypS,\parms}\,\prob{\parms}{\HypS}\,d\parms\,,
\end{equation}
which is often referred to as the \emph{marginal likelihood} (and sometimes as \emph{evidence}).

\subsection{Odds ratio and Bayes factor}
\label{sec:odds-ratio-bayes}

We can express the \emph{odds ratio} $\OSN$ between signal- and Gaussian-noise hypothesis for the
observed data $\dVx$ as
\begin{equation}
  \label{eq:9}
  \OSN(\dVx) \equiv \frac{\prob{\HypS}{\dVx}}{\prob{\HypN}{\dVx}}
  = \BSN(\dVx) \, \frac{\prob{\HypS}{\Inf}}{\prob{\HypN}{\Inf}} \,,
\end{equation}
where we used Bayes' theorem, namely
$\prob{a}{b}\prob{b}{\Inf}=\prob{b}{a}\prob{a}{\Inf}$,
to express $\OSN(\dVx)$ as a product of the prior hypothesis
odds and the so-called \emph{Bayes factor} $\BSN(\dVx)$, defined as
\begin{equation}
  \label{eq:10}
  \BSN(\dVx) \equiv \frac{\prob{\dVx}{\HypS}}{\prob{\dVx}{\HypN}}
  = \int_\parmSpace \Lr(\dVx;\parms)\, \prob{\parms}{\HypS}\,d\parms\,,
\end{equation}
where we used Eqs.~\eqref{eq:4},~\eqref{eq:7} and defined
the standard \emph{likelihood ratio} as
\begin{equation}
  \label{eq:11}
  \Lr(x;\parms) \equiv \frac{\prob{\dVx}{\HypS,\parms}}{\prob{\dVx}{\HypN}}
  = e^{\scalar{\dVx}{\dVh(\parms)} - \frac{1}{2}\scalar{\dVh(\parms)}{\dVh(\parms)}}\,.
\end{equation}
Inserting the transient-CW signal model \eqref{eq:5}, we can write this as
\begin{equation}
  \label{eq:12}
  \log \Lr(\dVx;\parms) = \Amp^\mu\, x'_\mu - \frac{1}{2}\Amp^\mu \,\M'_{\mu\nu}\,\Amp^\nu\,,
\end{equation}
where we defined
\begin{align}
  x'_\mu(\Dop,\Trans) \equiv \scalar{\dVx}{\dVh'_\mu}\,,   \label{eq:13a}\\
  \M'_{\mu\nu}(\Dop,\Trans) \equiv \scalar{\dVh'_\mu}{\dVh'_\nu}\,, \label{eq:13b}
\end{align}
generalizing the corresponding CW quantities, e.g.\ see \cite{2009CQGra..26t4013P}, to the transient signal
model.

\subsection{Maximum-likelihood: the $\F$-statistic}
\label{sec:freq-maxim-likel}

Contrary to the marginalization in Eq.~\eqref{eq:10} over unknown parameters $\parms$, which follows from the axioms
of probability, the orthodox ``maximum-likelihood'' approach consists of an \emph{ad-hoc} \emph{maximization}
of the likelihood ratio $\Lr$ over the unknown parameters $\parms$, i.e.\ we define the maximum-likelihood
statistic as
\begin{equation}
  \label{eq:15}
  \Lr_\ML(\dVx) \equiv \max_\parms \Lr(\dVx;\parms)\,.
\end{equation}
Given the explicitly quadratic dependency on $\Amp^\mu$ in Eq.~\eqref{eq:12}, this maximization can
be performed explicitly, which results in
\begin{equation}
  \label{eq:13}
  \ln \Lr_\ML(\dVx) = \max_{\{\Dop,\Trans\}} \F(\dVx;\Dop,\Trans)\,,
\end{equation}
were we encounter the well-known ``$\F$-statistic'', which was first derived in \cite{jks98:_data} for CW
signals. In the present transient-CW case, the transient $\F$-statistic is obtained explicitly as
\begin{equation}
  \label{eq:16}
  2\F(\dVx;\Dop,\Trans) \equiv x'_\mu\,\Minv'^{\mu\nu}\,x'_\nu\,,
\end{equation}
where $\Minv'^{\mu\nu}$ is defined as the inverse matrix of $\M'_{\mu\nu}$ of Eq.~\eqref{eq:13b}.

If the data $\dVx(t)$ contains a signal $\dVs(t;\parms_\sig)$, such that $\dVx = \dVn + \dVs(\parms_\sig)$,
then we can write Eq.~\eqref{eq:13a} as
\begin{equation}
  \label{eq:28}
  x'_\mu = n'_\mu + s'_\mu\,,
\end{equation}
with the obvious definitions $n'_\mu(\Dop,\Trans) \equiv \scalar{\dVn}{\dVh'_\mu}$ and
\begin{equation}
  \label{eq:34}
  s'_\mu(\parms_\sig; \Dop, \Trans) \equiv \scalar{\dVs(\parms_\sig)}{\dVh'_\mu(\Dop,\Trans)}\,,
\end{equation}
which depends both on the signal parameters $\parms_\sig = \{\Amp_\sig,\Dop_\sig,\Trans_\sig\}$ and the
matched-filter parameters $\{\Dop,\Trans\}$ of the ``template''.

Gaussian detector noise $\dVn(t)$ has zero mean, i.e.\ $\expect{\dVn}=0$, and therefore
$\expect{n'_\mu} = 0$ and $\expect{x'_\mu} = s'_\mu$. One can also show that the corresponding covariance is
$\expect{n'_\mu\,n'_\nu} = \M'_{\mu\nu}(\Dop,\Trans)$, and therefore
\begin{equation}
  \label{eq:30}
  \expect{x'_\mu\,x'_\nu} = \M'_{\mu\nu} + s'_\mu\,s'_\nu\,.
\end{equation}
Using this together with Eq.~\eqref{eq:16} one can further show that $2\F$ follows a $\chi^2$-distribution
with four degrees of freedom and non-centrality parameter $\rho^2$, i.e.
\begin{equation}
  \label{eq:32}
  \expect{2\F} = 4 + \rho^2\,,
\end{equation}
where the signal-to-noise ratio (SNR) $\rho$ is expressible as
\begin{equation}
  \label{eq:33}
  \rho^2(\parms_\sig; \Dop,\Trans) = s'_\mu\,\M'^{\mu\nu}\,s'_\nu\,.
\end{equation}
We see from Eq.~\eqref{eq:34} that the SNR will depend in a complicated way on the offset between signal
parameters $\parms_\sig$ and template parameters $\{\Dop,\Trans\}$ (see \cite{prix06:_searc} for the non-transient CW case).
In the special case of perfectly matched template parameters, i.e.\ $\Dop = \Dop_\sig$ and $\Trans = \Trans_\sig$,
we obtain the so-called ``optimal SNR'' $\rhoOpt$, which can be expressed as
\begin{equation}
  \label{eq:35}
  \rhoOpt^2(\parms_\sig) = \Amp_\sig^\mu\,\M'_{\mu\nu}(\Dop_\sig,\Trans_\sig)\;\Amp_\sig^\nu = \scalar{\dVs}{\dVs}\,.
\end{equation}

\subsection{Choice of signal priors}
\label{sec:choice-signal-priors}

In order to fully define the Bayes factor \eqref{eq:10}, we need to provide a complete signal hypothesis
including the prior probabilities $\prob{\parms}{\HypS}$ for the signal parameters $\parms = \{\Amp,\Dop,\Trans\}$.

\subsubsection{Prior on Doppler-parameters $\Dop$}
\label{sec:doppl-param-priors}

For simplicity we assume that the Doppler-parameters $\Dop$ are independent of amplitude- and
transient parameters $\{\Amp,\,\Trans\}$, so we can factor the full parameter prior $\prob{\parms}{\HypS}$ into
\begin{equation}
  \label{eq:68}
  \prob{\parms}{\HypS} = \prob{\Dop}{\HypS}\,\prob{\Amp,\Trans}{\HypS}\,,
\end{equation}
and so the Bayes factor \eqref{eq:10} now reads as
\begin{equation}
  \label{eq:69}
  \BSN(\dVx) = \int \BSN(\dVx;\,\Dop)\,\prob{\Dop}{\HypS}\, d^n\Dop\,,
\end{equation}
in terms of a ``targeted'' Bayes factor $\BSN(\dVx;\,\Dop)$ for a single Doppler point $\Dop$, namely
\begin{equation}
  \label{eq:70}
  \BSN(\dVx;\,\Dop) \equiv \int \Lr(\dVx;\Amp,\Trans,\Dop)\,\prob{\Amp,\Trans}{\HypS}\,d\Amp\,d\Trans\,.
\end{equation}
In the following we will focus exclusively on the \emph{targeted} Bayes factor, and sometimes drop $\Dop$
for simplicity of notation. The generalization to parameter searches over $\Dop$ is straightforward as given
by Eq.~\eqref{eq:69}.

\subsubsection{Prior on transient parameters  $\Trans$}
\label{sec:trans-wind-priors}

Astrophysically it would make sense to assume that the amplitude $h_0$ of a transient CW is related to its
timescale $\tTau$. For example, it might be reasonable to suspect that stronger transient GWs have
shorter duration and vice-versa, according to some prior on the total transient-CW energy emitted and on the distance
of such sources, cf.\ Sec.~\ref{sec:astr-toy-model}.
However, given the current lack of concrete astrophysical predictions to base such priors on, we assume a
simple independent prior $\prob{\Trans}{\HypS}$ for the transient signal parameters $\Trans$.
A naturally simple choice consists of independent uniform priors within some appropriate time-windows, i.e.\
\begin{equation}
  \label{eq:26}
  \begin{split}
    \prob{\tStart}{\HypS} &= \Uni{\tStart_{\min}}{\tStart_{\min}+\Delta\tStart}\,,\\
    \prob{\tTau}{\HypS}   &= \Uni{\tTau_{\min}}{\tTau_{\min}+\Delta\tTau}\,,
  \end{split}
\end{equation}
where we write uniform probability densities as $\Uni{a}{b} \equiv 1/(b - a)$ for the parameter falling inside
$[a,b]$ and zero otherwise.
We would also need priors on the window \emph{type}, e.g.\ $\prob{\type=\rect}{\HypS}=\prob{\type=\ex}{\HypS}=\frac{1}{2}$,
in order to marginalize over $\type$. However, for simplicity we will often assume a particular window-type $\type$ as
given and only marginalize over $\tStart, \tTau$. The effect of assuming an incorrect window-function within
$\type\in\{\rect,\ex\}$ is numerically studied in Sec.~\ref{sec:comp-diff-trans}, and appears to entail only
mild losses of detection power.

\subsubsection{Prior on amplitude-parameters $\Amp$}
\label{sec:ampl-param-priors}

Physically reasonable priors on the angle parameters $\{\cosi,\,\psi,\,\phi_0\}$ are relatively easy
to obtain if we assume ignorance about the orientation of the spinning neutron star (e.g.\ see
\cite{2009CQGra..26t4013P} for a more detailed discussion), namely by symmetry one can obtain
\begin{equation}
  \label{eq:21}
  \begin{split}
    \prob{\cosi}{\HypS} &= \Uni{-1}{1}\,,\\
    \prob{\psi}{\HypS}  &= \Uni{-\frac{\pi}{4}}{\frac{\pi}{4}}\,\\
    \prob{\phio}{\HypS} &= \Uni{0}{2\pi}\,.
  \end{split}
\end{equation}
The choice of prior for the overall amplitude parameter $h_0$ is less obvious: one could choose a scale-free
Jeffrey's prior, or a simpler uniform prior on some physically meaningful domain.

The downside of the isotropic amplitude priors \eqref{eq:21} is that the resulting marginalization over
$\Amp$ cannot be performed analytically, and computing the Bayes factor would require a numerical
integration over $\Amp$ in every point $\Dop,\Trans$, which will be computationally prohibitive.
However, as shown in \cite{2009CQGra..26t4013P}, by using an unphysical \emph{uniform prior} on the
4-\emph{vector} $\Amp^\mu$, one can \emph{analytically} marginalize over $\Amp$ and obtain a Bayes factor
\eqref{eq:70} expressed in terms of the well-known $\F$-statistic \eqref{eq:16}.

This choice has the major advantage of simplicity and computational efficiency, while incurring only small
losses in detection power compared to the more physical prior \eqref{eq:21}, as shown in \cite{2009CQGra..26t4013P}.
Given that fast and mature codes exist to compute $\F$ on real detector data (cf.\ \cite{prix:_cfsv2,lalsuite}),
we choose this prior as a convenient practical approximation.

However, as will be seen in the following, the original formulation of the uniform $\Amp^\mu$-prior in
\cite{2009CQGra..26t4013P} leads to a somewhat weak detection statistic for transient-CW signals and can be
improved by a minor modification.
The original ``constant-$\hmax$'' prior was defined as
\begin{equation}
  \label{eq:19}
  \prob{\Amp^\mu}{\HypS} = \left\{\begin{array}{ll}
      C \hspace*{0.5cm} & \mbox{if } h_0(\Amp) < \hmax\,,\\
      0 & \mbox{otherwise}\,,
    \end{array}\right.
\end{equation}
where $\hmax$ is a maximum cutoff amplitude needed in order to normalize the prior.

The $\Amp$-integration in Eq.~\eqref{eq:70} can be performed analytically with this prior, provided the data does
not cause the likelihood to peak close to the upper cutoff $\hmax$.
Namely, if the value of the integrand $\Lr(x;\Amp)$ is already negligible at the cutoff boundary, we can
extend the domain to infinity and obtain a 4-dimensional Gaussian integral, namely
\begin{equation}
  \label{eq:22}
  \int^{h_0<\hmax} \Lr(x;\Amp,\Trans)\,C\,d^4\!\Amp \approx \frac{(2\pi)^2\,C}{\sqrt{|\M|}}\,e^{\F(x;\Trans)}\,,
\end{equation}
where $|\M| \equiv \det\M$ is the determinant of the matrix $\M_{\mu\nu}$ of Eq.~\eqref{eq:13b}.
Using this approximation, we can therefore write the Bayes factor \eqref{eq:70} as
\begin{equation}
  \label{eq:23}
  \BSN(\dVx;\Dop) \approx \int \frac{(2\pi)^2 \,C}{|\M|^{1/2}}\,e^{\F(\dVx;\Trans)}\,\prob{\Trans}{\HypS}\,d\Trans\,.
\end{equation}
The Jacobian $J$ of the coordinate transformation $\Amp^\mu = \Amp^\mu(h_0,\cosi,\psi,\phio)$ is
$J = \frac{1}{4}{h_0^3}\left(1 - \cos^2\iota\right)^3$ (see \cite{2009CQGra..26t4013P}), and therefore
$d^4\!\Amp = J\,dh_0 \, d\!\cosi \; d\psi \, d\phio$.
We can now determine the normalization constant $C$ as
\begin{align}
  \label{eq:20}
  1 &= \int \prob{\Amp}{\HypS}\,d^4\!\Amp = C\, \frac{2\pi^2}{35}\,\hmax^{4}\,,
\end{align}
and obtain the constant-$\hmax$ Bayes factor explicitly as
\begin{equation}
  \label{eq:62}
  \BSNhmax = \frac{70}{\hmax^4\,\Delta\tStart\,\Delta\tTau}\int \frac{e^{\F}}{|\M|^{1/2}}\;d\tStart\,d\tTau\,,
\end{equation}
where we assumed a fixed window type $\type$.
The antenna-pattern weighting factor $|\M|^{-1/2}$ generally depends on the sky-position $\nhat$, the
transient-window type $\type$, start-time $\tStart$ and timescale $\tTau$, as seen from Eq.~\eqref{eq:13b}.
In the case of a fully targeted search with fixed $\tStart,\tTau$ and $\Dop$, as discussed in \cite{2009CQGra..26t4013P},
the weighting factor is constant and plays no role for the power of the detection statistic.
The fully targeted Bayes factor is therefore strictly equivalent to the detection power of the $\F$-statistic.

Interestingly, in the case of transient signals we find that the presence of this antenna-pattern weighting factor
in Eq.~\eqref{eq:62} seems to \emph{degrade} the detection power, and for some choices of parameter ranges
$\BSNhmax$ performs \emph{worse} than the maximum-likelihood statistic \eqref{eq:13}, namely
\begin{equation}
  \label{eq:27}
  {\Fmax}(x;\,\type) \equiv \max_{\{\tStart,\tTau\}} \F(x;\,\type,\tStart,\tTau)\,.
\end{equation}
However, a simple modification of the cutoff boundary in Eq.~\eqref{eq:19} allows us to eliminate the
antenna-pattern weighting factor $|\M|^{-1/2}$ in Eq.~\eqref{eq:23}.
Namely, by introducing an ``SNR-scale'' $\rhoh \equiv h_0\,|\M|^{1/8}$, and using a cutoff  $\rhoh < \rhohMax$
as the outer boundary of the domain instead of the amplitude-cutoff $h_0 < \hmax$, the modified \emph{ad-hoc}
``$\F$-statistic prior'' is now
\begin{equation}
\label{eq:63}
\prob{\Amp^\mu}{\HypS,\Trans} = \left\{\begin{array}{ll}
    \hat{C} \hspace*{0.5cm}& \mbox{if } \rhoh(\Amp) < \rhohMax\,,\\
    0 & \text{otherwise}\,,
    \end{array}\right.
\end{equation}
which results in the new transient Bayes factor
\begin{equation}
  \label{eq:25}
  \BF(\dVx;\,\Dop,\type) = \frac{70}{\rhohMax^4\, \Delta\tStart\,\Delta\tTau} \int e^{\F(\dVx;\Dop,\Trans)}\,d\tStart\,d\tTau\,,
\end{equation}
where in the following we denote $\BF\equiv\BSNrhohMax$as our detection statistic of choice.

As discussed in Sec.~\ref{sec:comp-sens-const}, numerical simulations show that this detection statistic is
more powerful than both $\BSNhmax$ as well as the orthodox maximum-likelihood statistic.
More work would be required to study these priors in more detail, and to develop a more physical choice of
priors that would be equally practical.

\subsection{SNR-loss due to rectangular-window offsets}
\label{sec:snr-loss-due}

Let us consider the effect of an offset in transient-window parameters $\Trans$ from the signal parameters
$\Trans_\sig$, assuming perfectly-matched Doppler parameters, i.e.\ $\Dop = \Dop_\sig$.
The dependence of the matched-filter SNR \eqref{eq:33} on
Doppler offsets $\Dop_\sig - \Dop$ signals has already been studied in great detail, e.g.\ see
\cite{brady98:_search_ligo_periodic,jones05:_ptole_metric,prix06:_searc}.

In the following we drop $\Dop$ and write Eq.~\eqref{eq:34} more explicitly as
\begin{equation}
  \label{eq:36}
  s'_\mu(\Amp_\sig,\Trans_\sig,\Trans) = \Amp^\nu_\sig\,\scalar{\dVh'_\nu(\Trans_\sig)}{\dVh'_\mu(\Trans)}\,,
\end{equation}
and using Eqs.~\eqref{eq:14} and \eqref{eq:inner_product}, we can further expand this as
\begin{equation}
  \label{eq:37}
  s'_\mu = \frac{2\Amp_\sig^\nu}{\Sn} \int h_\nu(t)\,h_\mu(t)\,\win_{\type_\sig}(t;\tStart_\sig,\tTau_\sig) \win_{\type}(t;\tStart,\tTau)\,d t\,,
\end{equation}
where we have omitted multi-detector summation for simplicity of notation.
It will be difficult to make further analytic progress with this expression in the general case, but we can
analyze the interesting special case of the rectangular window function \eqref{eq:2}, i.e.\ $\type_\sig=\type=\rect$.
In this case the window-functions simply truncate the integral, and so we obtain
\begin{equation}
  \label{eq:38}
  s'_\mu = \frac{2\Amp_\sig^\nu}{\Sn} \int_{t_0}^{t_1} h_\nu(t)\,h_\mu(t)\,d t\,,
\end{equation}
where we defined
$t_0 \equiv \max(\tStart_\sig,\tStart)$, and
$t_1 \equiv \min\left( (\tStart_\sig+\tTau_\sig), (\tStart+\tTau)\right)$.
Note that $[t_0,t_1]$ denotes the rectangular overlap between $\Trans_\sig$ and $\Trans$, and in the above
integral we assumed $t_1 \ge t_0$, otherwise the expression is zero.

\begin{figure}[htbp]
  \includegraphics[width=\columnwidth]{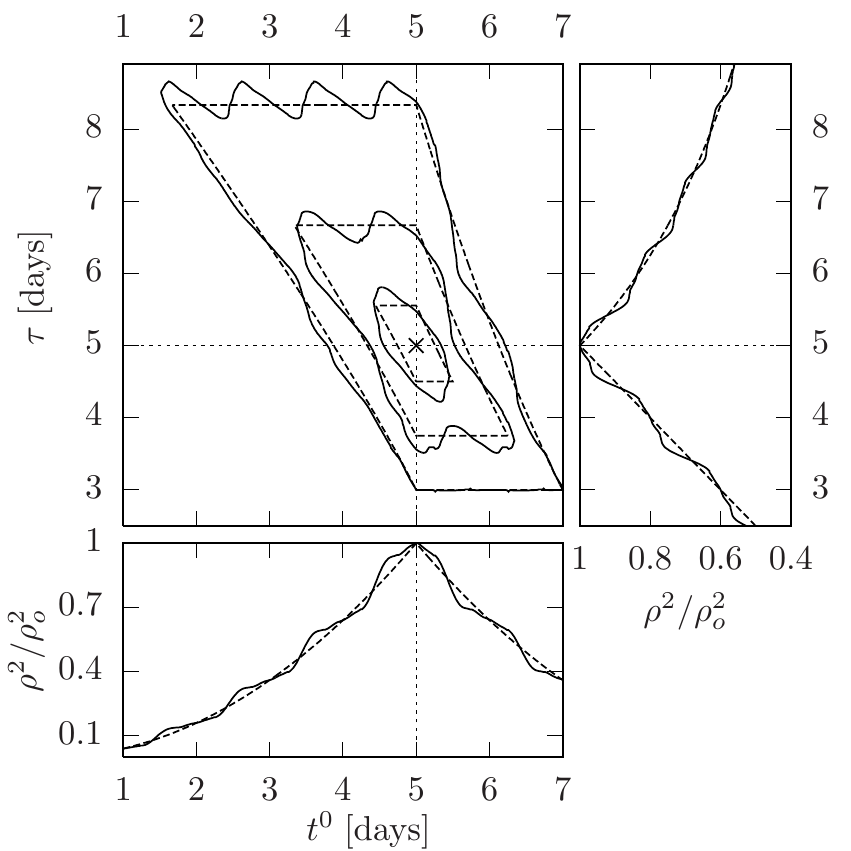}
  \caption{Approximate mismatch \eqref{eq:42} (dashed lines) and measured SNR loss (solid lines) as functions
    of the rectangular-window offsets $\{\tStart,\tTau\}$. The top left plot shows mismatch contour lines at
    $m=0.90, 0.75, 0.60$ respectively, and the side-panels show cross-sections of
    $\rho^2/\rhoOpt^2$ at fixed $\tStart=\tStart_\sig$ (right panel), and at fixed $\tTau=\tTau_\sig$ (bottom
    panel).
  }
  \label{fig:Mismatches}
\end{figure}

In order to simplify this even further, we note that the antenna-pattern matrix in Eq.~\eqref{eq:13b}
can be written as
\begin{equation}
  \label{eq:39}
  \M'_{\mu\nu} = \frac{2\tau}{\Sn}\,m_{\mu\nu}\,,
\end{equation}
where we defined $m_{\mu\nu} = \avg{h_\mu\,h_\nu}_\Sn$, in terms of the multi-detector time-average
$\avg{\ldots}_\Sn$  introduced in [Eq.~(59)] in \cite{prix06:_searc}.
The antenna-pattern functions are periodic with period of a sidereal day, and so the average $m_{\mu\nu}$
will be weakly oscillatory and converges to a constant for \mbox{$\tTau\gg1\mathrm{d}$}. Let us therefore approximate $m_{\mu\nu}$ as
constant for fixed $\Dop$, i.e.\ $m_{\mu\nu}\approx \bar{m}_{\mu\nu}$, which allows us to write Eq.~\eqref{eq:38} as
\begin{equation}
  \label{eq:40}
  s'_\mu \approx \frac{2\tTauDelta}{\Sn} \Amp_\sig^\nu\,\bar{m}_{\nu\mu}\,,\quad
  \text{with}\quad \tTauDelta \equiv \left[ t_1 - t_0 \right]^{+}\,,
\end{equation}
where $[\ldots]^+$ is the positivity operator, defined as $[x]^{+} = x$ for $x>0$ and zero otherwise.
Therefore $\tTauDelta$ is the length of overlap between signal and template windows.
Using $\M'_{\mu\nu} \approx (2\tTau/\Sn)\,\bar{m}_{\mu\nu}$, we obtain the mismatched SNR after
substituting into Eq.~\eqref{eq:33}, namely
\begin{equation}
  \label{eq:41}
  \rho^2 \approx \frac{\tTauDelta^2}{\tTau}\, \frac{2}{\Sn}\,\Amp_\sig^\mu\,\bar{m}_{\mu\nu}\,\Amp_\sig^\nu
  = \frac{\tTauDelta^2}{\tTau\,\tTau_\sig}\, \rhoOpt^2\,,
\end{equation}
in terms of the perfectly-matched ``optimal SNR'' $\rhoOpt$, defined in Eq.~\eqref{eq:35}.
Note that always $\tTauDelta \le \min(\tTau,\tTau_\sig)$, and equality only holds in the perfect-match case.
Expressing this in terms of the usual definition of mismatch $\mis$, we obtain
\begin{equation}
  \label{eq:42}
  \mis(\tStart_\sig,\tTau_\sig;\tStart,\tTau) \equiv \frac{\rhoOpt^2 - \rho^2}{\rhoOpt^2} \approx
  1 - \frac{\tTauDelta^2}{\tTau\,\tTau_\sig}\,.
\end{equation}
The behavior of this approximate mismatch function and the corresponding measured SNR loss is illustrated
in Fig.~\ref{fig:Mismatches} for a start-time $\tStart=5\,\days$ and duration $\tTau=5\,\days$
\footnote{All numerical results refer to a detector the Hanford site (``H1''), sky-location of the Crab pulsar,
  RA=05h34m31.973s, DEC=22:00:52.06, and start-time $\tStart=0$ with respect to a reference time of
  $\tRef=814838413\,\secs$ in GPS seconds (Nov 1, 2005).}.
Contrary to the well-known mismatch behavior in Doppler parameters $\Dop$, the transient mismatch metric
is not differentiable at the perfect-match point $\tTau = \tTau_\sig = \tTauDelta$, as seen in
Eq.~\eqref{eq:42} and Fig.~\ref{fig:Mismatches}, where the mismatch has a kink.
Therefore we cannot Taylor-expand the mismatch around the signal location and define a metric tensor
from the second-order quadratic form. Furthermore, the mismatch \eqref{eq:42} depends not only on the
parameter offsets, but also on the \emph{actual value} of the signal duration $\tTau_\sig$. Therefore the
mismatch function is not constant over the parameter space.
However, it is interesting to note that the mismatch behaves close to linearly around the point of perfect
match, and we can obtain the first-order variation of the mismatch \eqref{eq:42} as
\begin{equation}
  \label{eq:98}
  d\mis(\Trans_\sig) = 2\frac{|d\tStart|}{\tTau_\sig} + \frac{|d\tTau|}{\tTau_\sig}\,,
\end{equation}
which shows that the mismatch increases twice as fast for offsets in start-time $\tStart$ than for offsets
in timescale $\tTau$, which is also seen in Fig.~\ref{fig:Mismatches}. Also, as seen in
Fig.~\ref{fig:Mismatches}, the parameters $\tStart$ and $\tTau$ are correlated, and
the iso-mismatch curves close to the signal are straight lines with steepness $|d\tTau/d\tStart|=-2$, as seen from
Eq.~\eqref{eq:98}.

These properties are important for covering the transient-parameter space with a template bank, where one
tries to use the smallest number of templates while guaranteeing a certain minimum match for any signal within
the template-space.
For our present purpose it will be sufficient to make sure that the finite step-sizes in $\tStart,\tTau$ are
\emph{fine enough} to ensure a reasonable approximation to the integral~\eqref{eq:25}.
The worst-case mismatch \eqref{eq:42} occurs for the shortest timescale $\tTau_\sig$, so for
$\tTau_{\min}=0.5\,\days$, say, and a time-sampling in steps of $\TSFT = 1800\,\secs$, the worst case mismatch will
be bounded by $\mis \lesssim\TSFT/\tTau_{\min} \approx 4\%$.

\subsection{Semi-coherent Bayes factor}
\label{sec:semi-coher-trans}

Increasing the coherent integration time (or in our case, the maximal timescale $\tTau$) in wide parameter-space CW
searches over unknown Doppler parameters $\Dop$ typically results in a dramatic increase in computing cost.
The reason is that the likelihood function $\F$ becomes increasingly finely structured over Doppler parameters
$\Dop$, such that more and more templates need to be sampled in order to cover the parameter space (see
\cite{brady98:_search_ligo_periodic,jones05:_ptole_metric,prix06:_searc}).
This feature severely limits the feasible coherence time to $\sim\Ord{\days}$ for fully coherent searches over
unknown Doppler parameters.

The usual approach to this problem is to abandon fully coherent integration of Eq.~\eqref{eq:inner_product} over the
whole observation time and instead
adopt a ``semi-coherent'' approach: this is typically achieved by relaxing the constraint of a consistent
signal phase over the whole lifetime. Namely, one
splits the observation time $\Tobs$ into $\Nseg$ \emph{segments} of duration $\Tseg$, such that $\Tobs = \Nseg\,\Tseg$,
and requires phase-coherence only over each segment $\Tseg$. The initial phase $\phio$ is part of the set of four
amplitude-parameters $\Amp$, but for simplicity one relaxes the consistency-constraints for all four
amplitude-parameters $\Amp^\mu$ across different data-segments $\dVx_\iSeg$ with $i = 1\ldots\Nseg$.
This corresponds to replacing the four unknown amplitude parameters $\Amp^\mu$ by $\Nseg\times4$ unknown
amplitude parameters $\{\Amp^\mu_\iSeg\}_{i=1}^\Nseg$.
Using the product rule for joint probabilities of independent events, namely
$\prob{ \{\dVx_\iSeg\}}{\ldots} = \prod_{i=1}^\Nseg \prob{\dVx_\iSeg}{\ldots}$, we can express the
corresponding semi-coherent transient Bayes factor \eqref{eq:10} as
\begin{align}
  \label{eq:54}
  \BSN^\scoh&(\dVx) = \int d\Dop \int d\Trans \, \prob{\Dop,\Trans}{\HypS} \times \notag\\
  &\prod_{i=1}^\Nseg\int \Lr(\dVx_\iSeg;\Amp_\iSeg, \Dop,\Trans)\,\prob{\Amp_\iSeg}{\HypS} d^4\Amp_\iSeg \,.
\end{align}
Using the $\F$-statistic prior \eqref{eq:63}, we obtain
\begin{equation}
  \label{eq:56}
  \BF^\scoh(\dVx;\,\Dop,\type) = \frac{1}{\Delta\tStart\,\Delta\tTau}\left(\frac{70}{\rhohMax^4}\right)^{\Nseg}\int e^{\SF(\dVx;\Dop,\Trans)}\,d\tStart\,d\tTau\,,
\end{equation}
where we defined the semi-coherent sum $\SF$ as
\begin{equation}
  \label{eq:57}
  \SF(\dVx;\Dop,\Trans) \equiv \sum_{i=1}^{\Nseg} \F(\dVx_\iSeg;\Dop,\Trans)\,.
\end{equation}
The semi-coherent sum can be shown to require substantially fewer templates in parameter space for the same
amount of data analyzed e.g.\ see\cite{2000PhRvD..61h2001B,2010PhRvD..82d2002P}, and can in many cases achieve
a higher sensitivity for wide parameter-space CW searches at fixed computing cost.
In order to maximize the available computing power, such searches are performed on large computer
clusters and on Einstein@Home\footnote{\url{http://www.einsteinathome.org}}, a large public distributed
computing project (e.g.\ see \cite{2009PhRvD..80d2003A} for an Einstein@Home search on LIGO data).

Therefore the semi-coherent Bayes factor \eqref{eq:56} allows for an efficient and computationally feasible
transient-CW search over a wide Doppler parameter space $\Dop$ of sources with unknown sky-position, frequency and
spindown. More work is required to fully develop and study this approach, as our present analysis is mostly
focused on the coherent method. A semi-coherent search for transient CWs would be very suitable to be
performed on the Einstein@Home computing platform.

\section{Detection efficiency}
\label{sec:detection-efficiency}

In order to quantify the detection efficiency of different statistics we perform Monte-Carlo studies with
simulated Gaussian noise and injected signals with parameters drawn according to their priors.
Note that in order to implement this in the most efficient way, we do not perform end-to-end simulations from
generated artificial data processed through the whole pipeline. Rather, we more directly synthesize draws of
the different statistics from their known distributions, or compute them from draws of intermediate
data-products. This is possible because we know the distribution of intermediate data-products in the case of
Gaussian noise. The details of this procedure are described in Appendix~\ref{sec:synth-monte-carlo}.

A useful method to compare different detection statistics is to compare their
``receiver-operator-characteristic'' (ROC), namely the expected detection probability $\probDet$ versus the
false-alarm probability $\probfA$ for some signal population. This follows the spirit of the
Neyman-Pearson criterion, which compares the detection probability of different statistics at fixed
false-alarm probability.

Note, however, that the traditional way of plotting ROC curves, namely $\probDet$ versus $\probfA$ is somewhat
unfortunate. We know \emph{a priori} that any statistic of positive detection power satisfies $\probDet > \probfA$,
and $\probDet = \probfA$ corresponds to a complete random guess. Therefore only the upper triangle of the plot
is of any interest, and it is more informative to plot $\probDet - \probfA$ versus $\probfA$ instead, which
quantifies how much better any statistic performs compared to a random guess.

\subsection{Comparing different amplitude priors}
\label{sec:comp-sens-const}

We first consider the detection efficiency of the two Bayes factors $\BSNrhohMax$ of Eq.~\eqref{eq:25}, and
$\BSNhmax$ of Eq.~\eqref{eq:62}. These only differ by their cutoff boundary on the uniform-$\Amp^\mu$
amplitude priors, as discussed in Sec.~\ref{sec:ampl-param-priors}.
For simplicity we consistently used a rectangular transient window \eqref{eq:2} for both the injections and
the search.
We performed $N=10^4$ random draws in the noise- and in the signal-case. We estimated the errors on
$\probDet$ using a jackknife estimator on 100 subsets, the resulting estimated 1$\sigma$ errors in the
following ROC curves are always less than $d(\probDet) \lesssim 0.02$.

The amplitude parameters are drawn according to their \emph{physical priors} \eqref{eq:21}. For the first
simulation we fixed the optimal SNR \eqref{eq:35} to $\rhoOpt=3$.
This is achieved by re-scaling $h_0$ according to the resulting SNR for drawn values of $\{\cosi,\psi,\phio\}$.
We always use the same ranges for the search and for drawing signal parameters from, and we consider two sets
of transient window parameter ranges, namely
\begin{equation}
  \label{eq:99}
  \begin{split}
    \text{\rangeI:}  \qquad & \tTau \in [0.1,0.6]~\days, \quad \tStart \in [0,9]~\days\,,\\
    \text{\rangeII:} \qquad & \tTau \in [0.5,2.5]~\days, \quad \tStart \in [0,6]~\days\,.
  \end{split}
\end{equation}

\begin{figure}[htbp]
  \centering
  \includegraphics[width=\columnwidth]{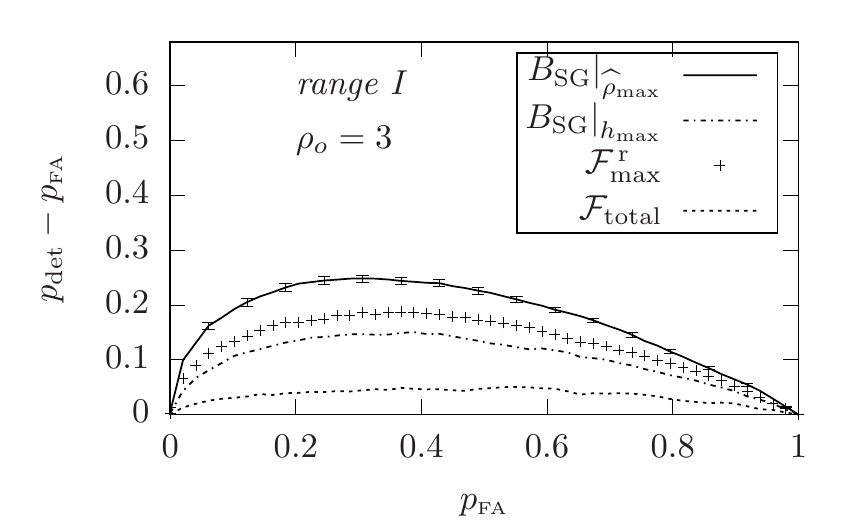}\\
  \includegraphics[width=\columnwidth]{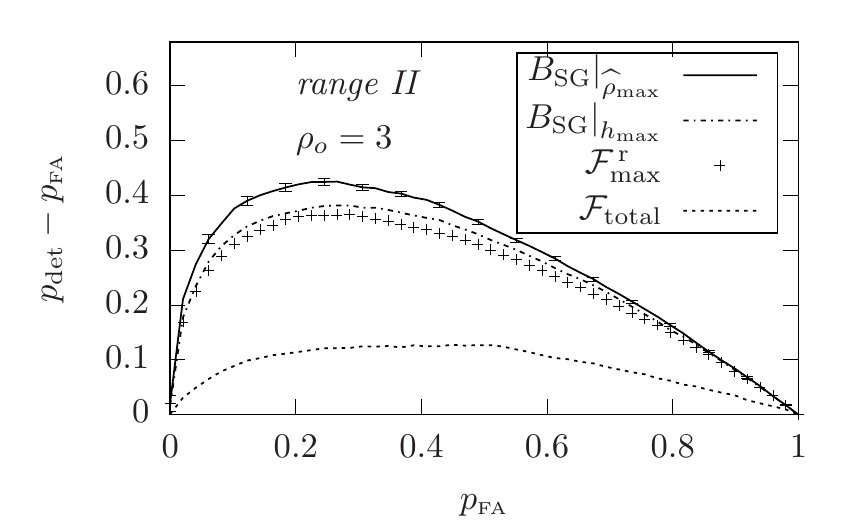}
  \caption{Detection efficiency of $\BSN|_{\rhohMax}$, $\BSN|_{\hmax}$ and $\Fmax$ for injected signals at
    fixed optimal SNR of $\rhoOpt=3$, using a rectangular transient window ($\type_\sig=\rect$).
    The curve labelled '$\F_{\mathrm{total}}$' refers to a standard CW $\F$-statistic search over the full data span.
    The upper plot corresponds to transient-window parameters drawn from \rangeI{}, while the lower plot is
    for \rangeII, see Eq.~\eqref{eq:99}.
    We only plot 1$\sigma$ error-bars for $\BSN|_{\rhohMax}$, which are representative for the size of the
    errors on the other curves.
  }
  \label{fig:ROC_BSN_reg}
\end{figure}

The results of this Monte-Carlo simulation are shown in Fig.~\ref{fig:ROC_BSN_reg}, and we see that the
statistic $\BF \equiv \BSNrhohMax$ seems to generally perform better than $\BSNhmax$, as discussed in
Sec.~\ref{sec:ampl-param-priors}.
For some choices of transient window ranges (such as \rangeI), the latter actually performs worse
than the orthodox maximum-likelihood statistic $\Fmax$, as seen in the upper plot in Fig.~\ref{fig:ROC_BSN_reg}.
We also note that the detection probability for signals of equal optimal SNR $\rhoOpt$ is lower for \rangeI{} than for \rangeII. This can be
understood from the substantially larger parameter space associated with \rangeI,
namely $\tTau_{\max}=0.6\,\days$ in $\Tobs=9\,\days$, as compared to $\tTau_{\max}=2.5\,\days$ in $\Tobs=6\,\days$ for \rangeII.
Therefore we can fit $\Tobs/\tTau=15$ independent rectangular windows into the observation time in
\rangeI, while for \rangeII{} this factor is only $2.4$. This entails more independent trials and therefore a
higher false-alarm probability for \rangeI{}, which reduces the detection power for signals of the same SNR.

\begin{figure}[htbp]
  \centering
  \includegraphics[width=\columnwidth]{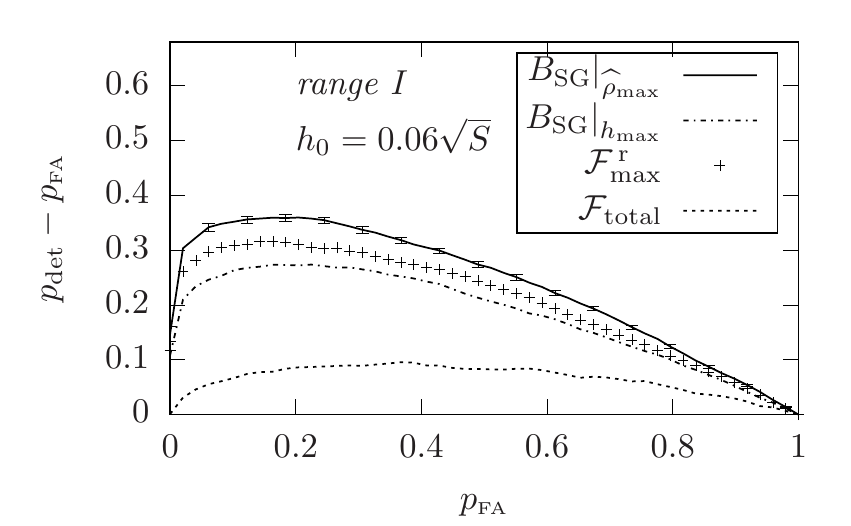}\\
  \includegraphics[width=\columnwidth]{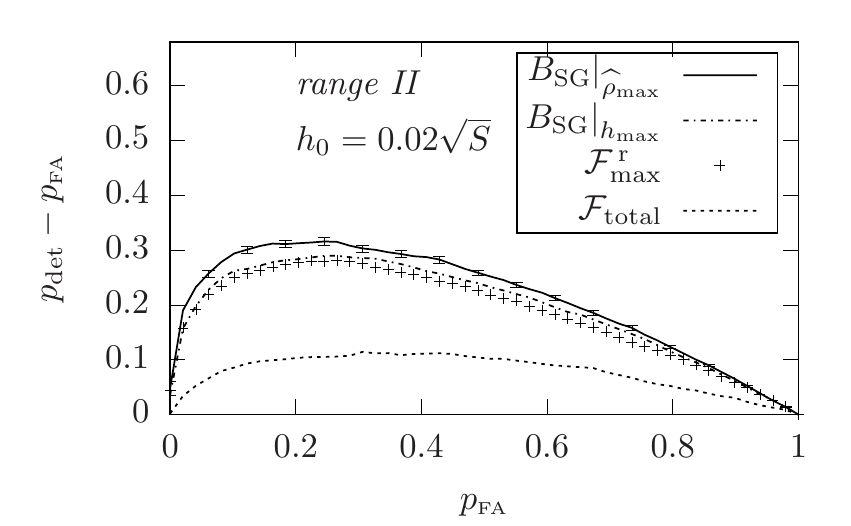}
  \caption{Same as Fig.~\ref{fig:ROC_BSN_reg}, with signals injected at fixed amplitude
    $h_0=0.06\sqrt{\Sn}$ and $h_0=0.02\sqrt{\Sn}$, respectively.}
  \label{fig:ROC_BSN_reg_fixedh0Nat}
\end{figure}

In order to verify that these qualitative conclusions are not restricted to injections at constant
SNR $\rhoOpt$, we also performed these Monte-Carlo simulations for injected signals at constant amplitude
$h_0/\sqrt{\Sn}$. Note that this results in a wide range of injected signal SNRs $\rhoOpt$ due to the varying signal durations
$\tTau$. The results of these simulations are shown in Fig.~\ref{fig:ROC_BSN_reg_fixedh0Nat}, which
qualitatively agree with Fig.~\ref{fig:ROC_BSN_reg}.

As expected, these results also show that the transient statistics are substantially more sensitive to
transient-CW signals than a standard ``infinite-duration'' CW $\F$-statistic search over the full span $\Tobs$.
The quantitative advantage in recovered SNR depends on the details of the transient parameter space, as seen
from the mismatch \eqref{eq:42}, namely setting $\tTau = \Tobs$ and $\tTauDelta = \tTau_\sig$, we find the recovered
fraction of $\rhoOpt^2$ is roughly proportional to the ``duty cycle'' $\tTau_\sig/\Tobs$ of the transient
signal with respect to the observation time $\Tobs$.

\subsection{Comparing rectangular and exponential windows}
\label{sec:comp-diff-trans}

Another question of interest is how robust the detection statistic $\BF^\type$ is, which assumes a particular
transient-window type $\type$, if the transient signal actually has a different window type $\type_\sig$.
We cannot answer this question in general, but it is instructive to study the simple case of
type-mismatch between the rectangular and exponential transient-window types.
We inject signals with either rectangular (Eq.~\eqref{eq:2}) or exponential (Eq.~\eqref{eq:3}) transient window, and
we perform the search using both rectangular and exponential transient windows, respectively.
The Monte-Carlo parameters are the same as in the previous section, and we only show the results for the
transient parameters \rangeII{} of Eq.~\eqref{eq:99}, the qualitative conclusions are the same for \rangeI.
\begin{figure}[htbp]
  \centering
  \includegraphics[width=\columnwidth]{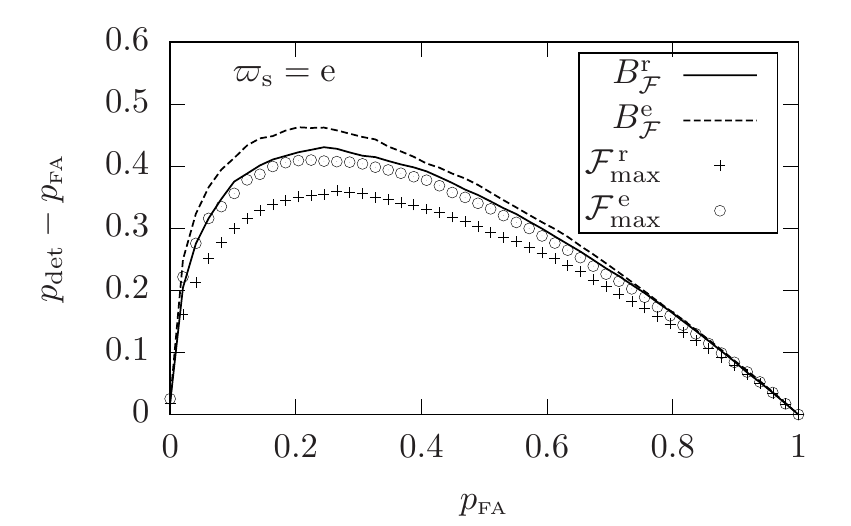}\\
  \includegraphics[width=\columnwidth]{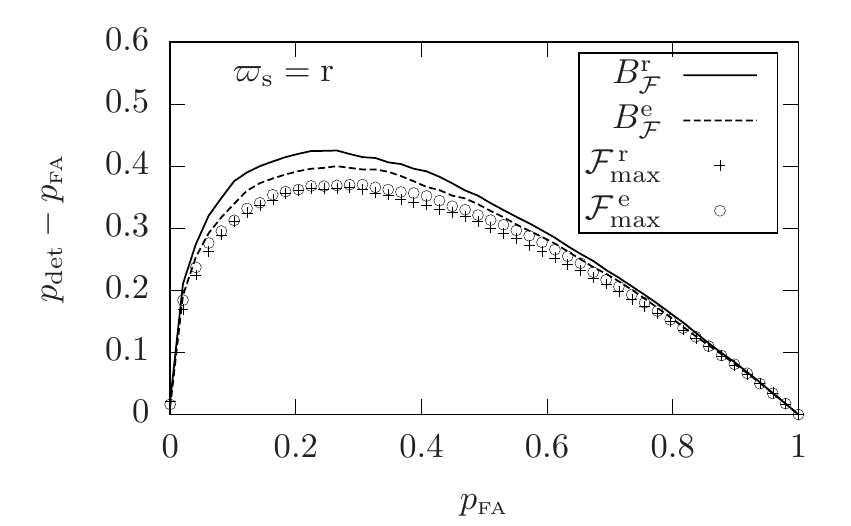}
  \caption{Detection efficiency for different transient-window \emph{types}. The plot shows detection power of
    the Bayes factor $\BF^\type$ and maximum-likelihood statistic $\Fmax^\type$ for assumed window-type
    $\type$ in the case of injected signals of window type $\type_\sig$.
    Transient parameters were drawn from \rangeII{} of Eq.~\eqref{eq:99} with a fixed SNR of $\rho=3$.
    The upper plot is for \emph{rectangular} injected transients ($\type_\sig = \rect$), while the lower plot
    is for exponential injected transients ($\type_\sig = \ex$).
  }
  \label{fig:compare_windows}
\end{figure}
The results of these Monte-Carlo simulations are shown in Fig.~\ref{fig:compare_windows}. We see that using
the wrong window-type, i.e.\ $\type \not= \type_\sig$, results in a loss of detection power in $\BF$, as  expected.
However, the loss is quite moderate, which in practice would favor a search assuming the simpler and much more
computationally efficient rectangular window type, $\type =\rect$ (cf.\ Sec.~\ref{sec:CompCost}), if the
search is computationally limited. If computing cost is not an issue, for example in a targeted search for
known pulsars, one could perform both searches and then marginalize over $\type$.
More importantly, however, these results suggest that we can hope to be reasonably sensitive also to
transient signals with \emph{different} time evolutions not contained in $\type\in\{\rect,\ex\}$.

It is interesting to note in Fig.~\ref{fig:compare_windows} that in the case of injecting rectangular-window
signals, i.e.\ $\type_\sig=\rect$, the maximum-likelihood $\Fmax^{\,\ex}$ assuming an exponential window, can
outperform the more correct $\Fmax^{\,\rect}$-statistic. A similar effect is seen in the corresponding
simulations for \rangeI, which are not shown here.
The origin of this ``anomaly'' can be traced to the fact that for the same timescale parameter $\tTau$, the
exponential-window waveform \eqref{eq:3} lasts three times longer than the rectangular-window waveform \eqref{eq:2}.
Therefore, the parameter spaces \rangeI{} and \rangeII{} contain substantially more independent trials for the
rectangular-window waveform compared to the exponential one. This results in lower false-alarm probabilities
at fixed threshold for  $\Fmax^{\,\ex}$ compared to $\Fmax^{\,\rect}$. Although there is a loss in SNR due to
window-type mismatch, this effect is partly weaker than the difference in false-alarm probabilities, resulting
in a partly more powerful detection statistic.
To test this explanation, we have repeated the Monte-Carlo simulation with empty ranges (i.e.\ a fixed
transient window), and with ranges where $\tTau_{\min} > \Delta\tStart$, such that all waveforms overlap in the
transient start-time range, independently of window-function type. In both cases the ``anomaly'' disappears.
Ranges I and II both have the feature that $\tTau_{\min} < \Delta\tStart$, such that the waveform-overlap will
be substantially lower for the rectangular-window compared to the exponential-window waveforms, resulting in a
large difference in independent trials in the noise case. These observations are consistent with the
explanation that the ``anomaly'' is caused by the difference in independent trials.
Interestingly, however, this effect is not observed for the marginalized Bayes-factors, which always seem to
correctly take into account the effective size of the parameter space.

\subsection{Bayes factor self-consistency condition}
\label{sec:odds-ratio-cons}

One can derive an interesting self-consistency condition from the general definition \eqref{eq:10} of the
Bayes factor, which provides a useful Monte-Carlo test of our implementation:
the probability of obtaining a Bayes factor $\BSN(\dVx) \in [B_0,\,B_0+d B]$ under any hypothesis
$\Hyp_i$ is given by the probability of obtaining a measurement $\dVx$ in the infinitesimal volume slice
$\Delta\Vol_0 \equiv \{\dVx \;:\; \BSN(\dVx) \in [B_0,\,B_0+dB]\}$ under that hypothesis, i.e.\
\begin{equation}
  \label{eq:64}
  \prob{B_0}{\Hyp_i}\,dB = \int_{\Delta\Vol_0} \prob{\dVx}{\Hyp_i}\, d^n\dVx\,.
\end{equation}
Changing local coordinates from $\dVx$ to $\detVec{y} = \{B(\dVx), \detVec{y}_\perp\}$,
where $y_\perp$ denotes $n-1$ coordinates on the constant-$B_0$ hypersurface
$\Surf_0 \equiv \{ \dVx \;:\; \BSN(\dVx) = B_0\}$, this can be written as
\begin{equation}
  \label{eq:66}
  \prob{B_0}{\Hyp_i} = \int_{\Surf_0} \prob{\dVx}{\Hyp_i}\,d\Surf\,,
\end{equation}
where we defined the surface element $d\Surf \equiv J\, d^{n-1} y_\perp$ and
we assume that the Jacobian $J$ is non-singular, i.e.\ $J\equiv|\partial\detVec{y}/\partial\dVx| \not=0$ everywhere
inside $\Delta\Vol_0$. Using the definition \eqref{eq:10} of the Bayes factor, we can substitute
$\prob{\dVx}{\HypS} = \BSN(\dVx)\, \prob{\dVx}{\HypN}$, and obtain
\begin{align}
  \label{eq:65}
  \prob{B_0}{\HypS} &= \int_{\Surf_0} \BSN(\dVx)\,\prob{\dVx}{\HypN}\,d\Surf \notag\\
  &= B_0 \, \int_{\Surf_0}\,\prob{\dVx}{\HypN}\,d\Surf\notag\\
  &= B_0 \,\prob{B_0}{\HypN}\,,
\end{align}
where we used the fact that $\left.\BSN(\dVx)\right|_{\Surf_0}=B_0$ by definition of $\Surf_0$.
We therefore find a general self-consistency relation for any Bayes factor, namely
\begin{equation}
  \label{eq:46}
  \BSN(\dVx) \equiv \frac{\prob{\dVx}{\HypS}}{\prob{\dVx}{\HypN}} = \frac{\prob{\BSN}{\HypS}}{\prob{\BSN}{\HypN}}\,.
\end{equation}
In the framework of Monte-Carlo simulations \cite{2008arXiv0804.1161S}, this implies that if we draw random
data $\dVx$ according to the priors assumed in the Bayes factor, the ratio of probability-densities of obtaining
$\BSN=B_0$ in the signal- and the noise-case is identical to $B_0$.
If the assumptions are satisfied, the signal-distribution of $\BSN$ is therefore not
independent of the noise-distribution, but is uniquely determined by it. If we know
$\prob{\BSN}{\HypN}$, then we also known the signal distribution, and vice versa.

The resulting self-consistency relation for the odds ratio \eqref{eq:9} is
\begin{equation}
  \label{eq:47}
  \OSN = \frac{\prob{\OSN}{\HypS}}{\prob{\OSN}{\HypN}}\,\frac{\prob{\HypS}{\Inf}}{\prob{\HypN}{\Inf}}\,,
\end{equation}
where the prior odds ratio determines the probability of drawing a sample $\dVx$ from the signal- or
noise-population, respectively. Therefore the odds ratio predicts the ratio of event densities at any
value $\OSN$, rather than the ratio of normalized probability densities.

Note that in order for Eq.~\eqref{eq:46} to hold for the transient-CW Bayes factor $\BF$ defined in
Eq.~\eqref{eq:25}, one must not draw signal amplitude parameters $\Amp^\mu$ according to the physical priors \eqref{eq:21}, but
according to the (unphysical) $\F$-statistic priors \eqref{eq:63} that went into the construction of $\BF$.
\begin{figure*}[htbp]
  \centering
  \mbox{
    \includegraphics[width=0.5\textwidth]{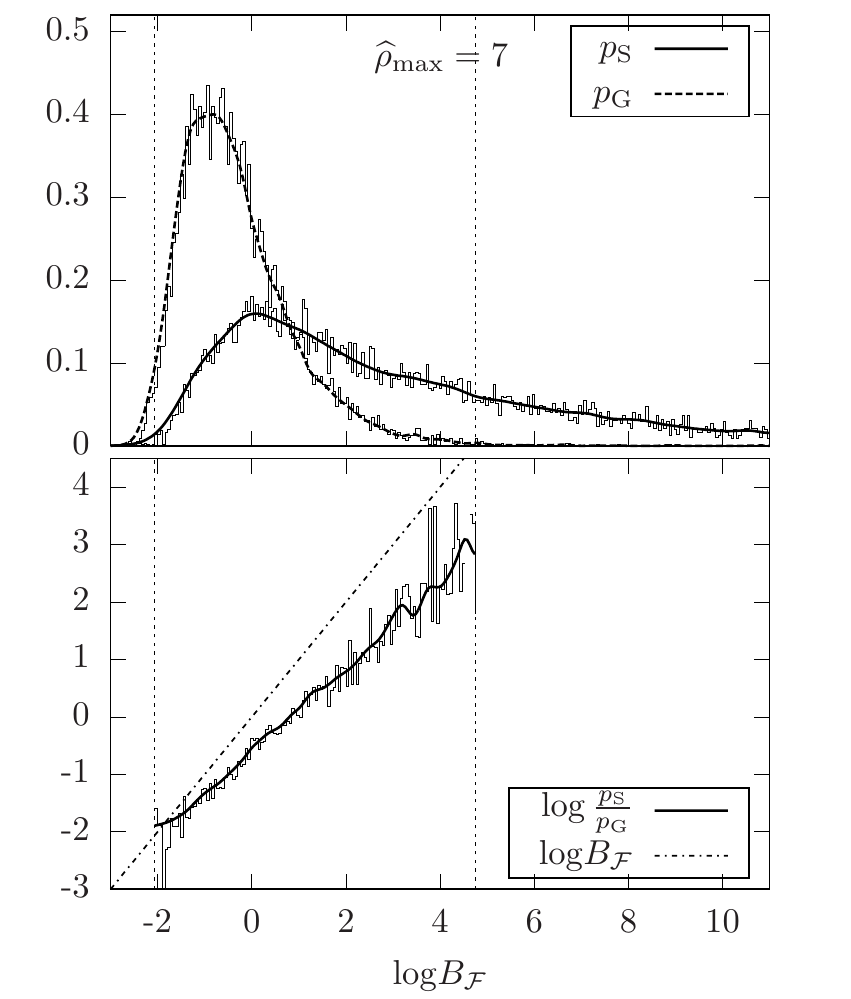}\hspace*{-1cm}
    \includegraphics[width=0.5\textwidth]{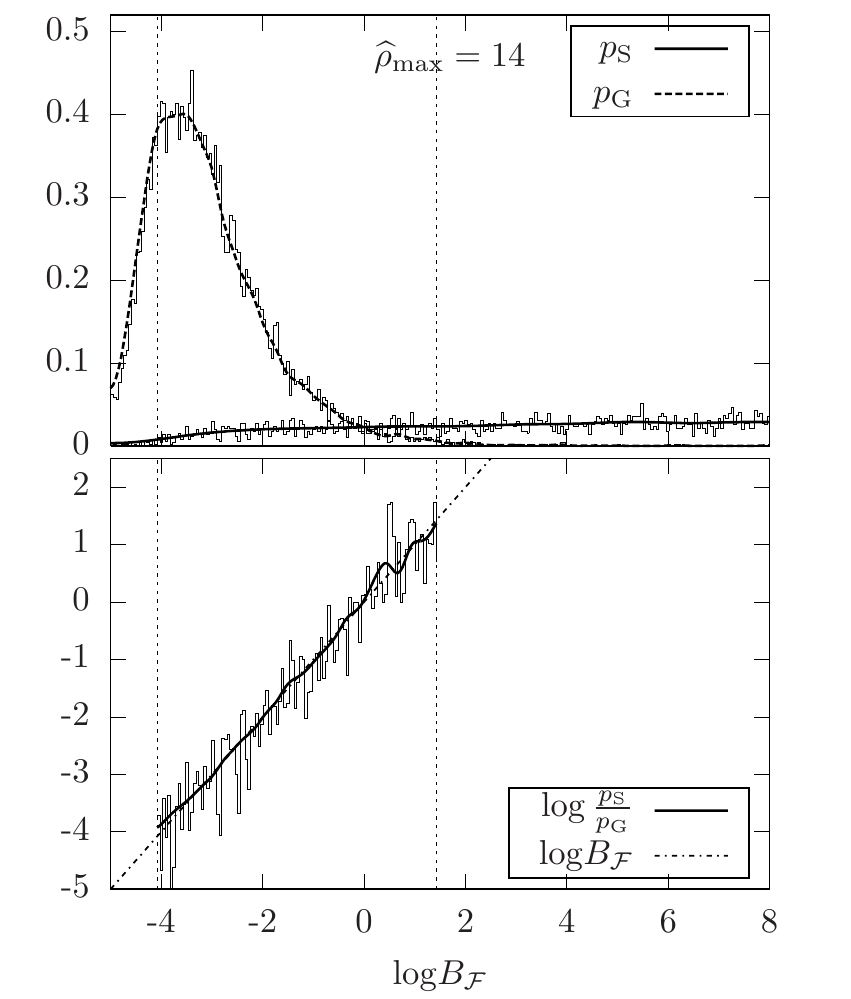}
  }
  \caption{Results of Monte-Carlo simulation of the Bayes-factor consistency relation \eqref{eq:100}, using
    $N=10^4$ draws, transient \rangeII, and amplitude priors \eqref{eq:63} with cutoff $\rhohMax=7$ (\emph{left panel}),
    and $\rhohMax=14$ (\emph{right panel}), respectively.
    The upper plots show the distributions of $\log\BF$ in the noise-case (dashed), and in the signal
    case (solid), and the lower plots show $\log(p_\Signal/p_\Noise)$, which should coincide with
    $\log\BF$ (dot-dashes line) according to \eqref{eq:100}.
    We show the binned histogram values (thin lines) as well as kernel-smoothed fits (thick lines), and we
    have restricted the consistency test to a region of good overlap (indicated by vertical dashed lines)
    between  both distributions in order to avoid numerical problems.
  }
  \label{fig:OSN_consistency}
\end{figure*}
The self-consistency relation \eqref{eq:46} can equivalently be expressed as
\begin{equation}
  \label{eq:48}
  \log\BF = \log \prob{\log\BF}{\HypS} - \log \prob{\log\BF}{\HypN}\,,
\end{equation}
which is more directly suitable for numerically testing this relation in a Monte-Carlo simulation.
Defining the shortcut $p_i \equiv \prob{\log\BF}{\Hyp_i}$, this can also be written as
\begin{equation}
  \label{eq:100}
  \log\frac{p_\Signal}{p_\Noise} = \log\BF\,.
\end{equation}
We have performed a Monte-Carlo simulation generating values of $\BF$ in the noise- and signal-cases, with
amplitude parameters $\Amp^\mu$ drawn according to the (unphysical) $\F$-statistic priors \eqref{eq:63}.
Figure~\ref{fig:OSN_consistency} shows the resulting distributions of $p_\Signal$ and $p_\Noise$, and the
plots of $\log (p_\Signal/p_\Noise)$ versus $\log\BF$, which should fall on a straight line of unit slope according to the
self-consistency relation \eqref{eq:100}.
These results show that the self-consistency relation is increasingly well satisfied with increasing prior
cutoff $\rhohMax$, in particular we find good agreement for cutoff values above $\rhohMax \gtrsim 10$, as illustrated in
Fig.~\ref{fig:OSN_consistency}.
This can be understood as follows: for smaller values of $\rhohMax$, the noise population $\prob{\BF}{\HypN}$ is
biased towards larger values of $\BF$, because the approximation in Eq.~\eqref{eq:22} is increasingly violated.
In the noise-only case, the likelihood ratio $\Lr$ will peak somewhere around $\Amp^\mu=0$ and fall
off according to a Gaussian \eqref{eq:12} with characteristic width of order
$\sigma(\rhoh)\sim \Ord{1}$, modulo geometric factors or order unity. Therefore the value of $\Lr$ will \emph{not} be
negligible at the cutoff boundary $\rhohMax \sim \Ord{1}$, and the extension to infinity will
\emph{overestimate} the integral. Therefore $\rhohMax\gg 1$ is necessary for the integral to be well
approximated by Eq.~\eqref{eq:22}.

\section{Parameter estimation}
\label{sec:parameter-estimation}

Parameter estimation simply consists of computing the posterior probability for the signal parameters
$\parms$, given the observed data $\dVx$, namely
\begin{equation}
  \label{eq:49}
  \prob{\parms}{\dVx,\HypS} = c\,\prob{\dVx}{\HypS,\parms}\,\prob{\parms}{\HypS}\,,
\end{equation}
where $c = 1/\prob{\dVx}{\HypS}$ is a normalization constant independent of $\parms$.
\begin{figure*}[tbp]
  \centering
  \mbox{
    \includegraphics[width=\columnwidth]{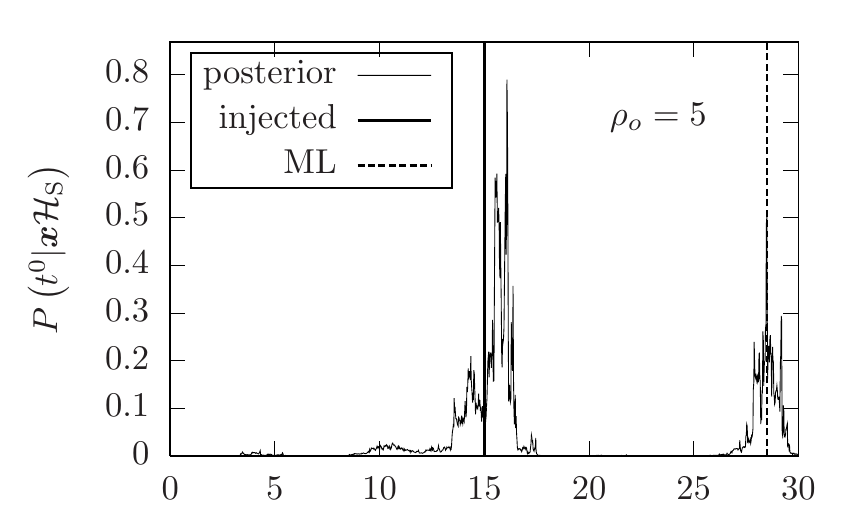}
    \includegraphics[width=\columnwidth]{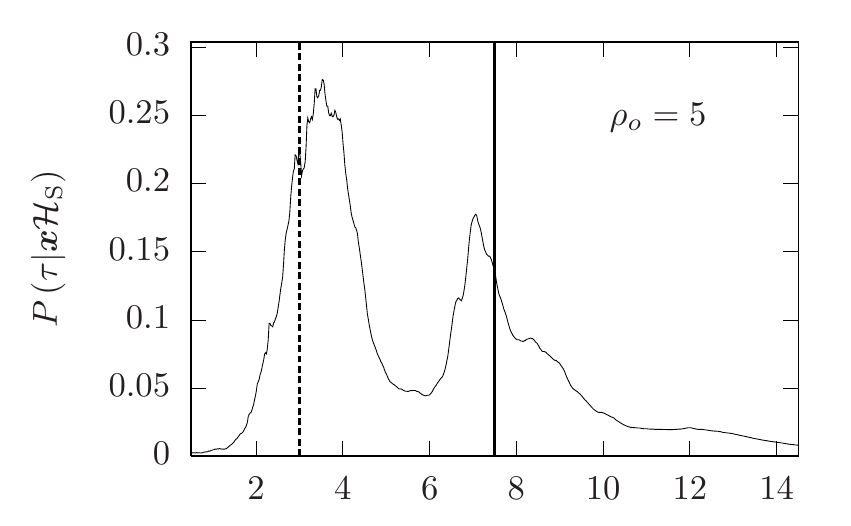}
    }\\[-0.5cm]
  \mbox{
    \includegraphics[width=\columnwidth]{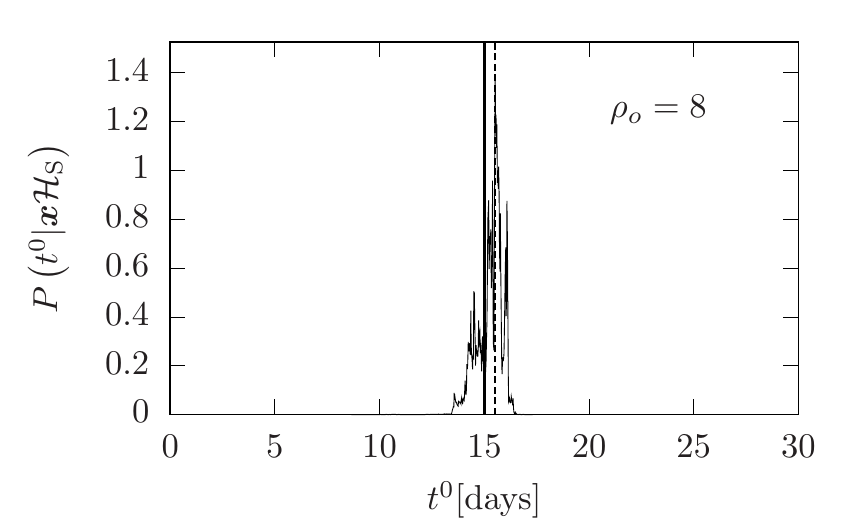}
    \includegraphics[width=\columnwidth]{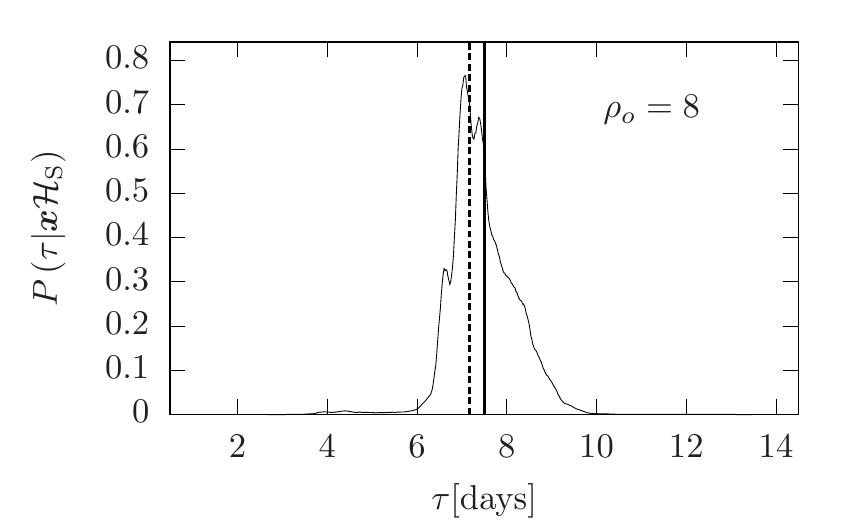}
    }
  \caption{Example posteriors \eqref{eq:53} on $\tStart$ (left column) and $\tTau$ (right column), for an
    injected rectangular transient-CW signal within \rangeIII{}, with randomly drawn amplitude parameters at
    fixed optimal SNR $\rhoOpt=5$ (upper row) and $\rhoOpt=8$ (lower row).
    The solid vertical line indicates the injected parameter value, and the dashed vertical line ('ML')
    indicates the maximum-likelihood estimate.}
  \label{fig:paramEstimation_pdfs}
\end{figure*}
Often one is not interested in a simultaneous estimate of the full set of parameters $\parms$, but only in a
subset $\parms_1\subset\parms$ without regard for the ``nuisance'' parameters $\parms_2$,
where $\parms = \{\parms_1, \parms_2\}$. From the general expression
$\prob{\parms_1}{\ldots} = \int\prob{\parms_1,\parms_2}{\ldots}\,d\parms_2$ and Bayes' theorem
\eqref{eq:49}, we obtain the marginalized posterior
\begin{equation}
  \label{eq:50}
  \prob{\parms_1}{\dVx,\HypS} \propto \int \prob{\dVx}{\HypS,\parms_1,\parms_2}\,\prob{\parms_1,\parms_2}{\HypS}\,d\parms_2\,.
\end{equation}
For our present model the likelihood function \eqref{eq:7} can be expressed as
$\prob{\dVx}{\parms} = \kappa e^{-\scalar{\dVx}{\dVx}/2}\,\Lr(x;\parms)$ in terms of the likelihood ratio
$\Lr$ of Eq.~\eqref{eq:12}. Using $\F$-statistic priors \eqref{eq:63}, we can perform the
$\Amp^\mu$-integration explicitly and obtain
\begin{equation}
  \label{eq:51}
  \prob{\Trans,\Dop}{\dVx,\HypS} \propto e^{\F(\dVx;\Dop,\Trans)}\,\prob{\Trans,\Dop}{\HypS}\,,
\end{equation}
which is a useful starting point for further marginalization.
If we consider a targeted search in Doppler parameter,
i.e.\ $\prob{\Dop}{\HypS} = \delta(\Dop - \Dop_\sig)$, with an assumed window function type $\type$, we can
write the posterior probability for $\{\tStart, \tTau\}$ as
\begin{equation}
  \label{eq:52}
  \prob{\tStart,\tTau}{\dVx,\HypS,\Dop,\type} \propto e^{\F(\dVx;\Dop,\Trans)}\,\prob{\tStart,\tTau}{\HypS}\,,
\end{equation}
and the respective marginal posteriors on the transient parameters are simply
\begin{equation}
  \label{eq:53}
  \begin{split}
    \prob{\tStart}{\dVx,\HypS,\Dop,\type} &\propto \int e^{\F(\dVx;\Dop,\Trans)}\,d\tTau\,,\\
    \prob{\tTau}{\dVx,\HypS,\Dop,\type} &\propto \int e^{\F(\dVx;\Dop,\Trans)}\,d\tStart\,,
  \end{split}
\end{equation}
where we assumed uniform priors \eqref{eq:26} for $\{\tStart,\tTau\}$. The generalization to marginalization
over the window type $\type$ is straightforward, and yields a weighted sum of these posteriors with relative
prior probabilities of the different window types, e.g.\
\begin{equation}
  \label{eq:24}
  \prob{\tStart}{\dVx,\HypS,\Dop} \propto \sum_{\type} \prob{\tStart}{\dVx,\HypS,\Dop,\type}\,\prob{\type}{\HypS}\,.
\end{equation}
\begin{figure*}[htbp]
  \centering
  \mbox{
    \includegraphics[width=0.5\textwidth]{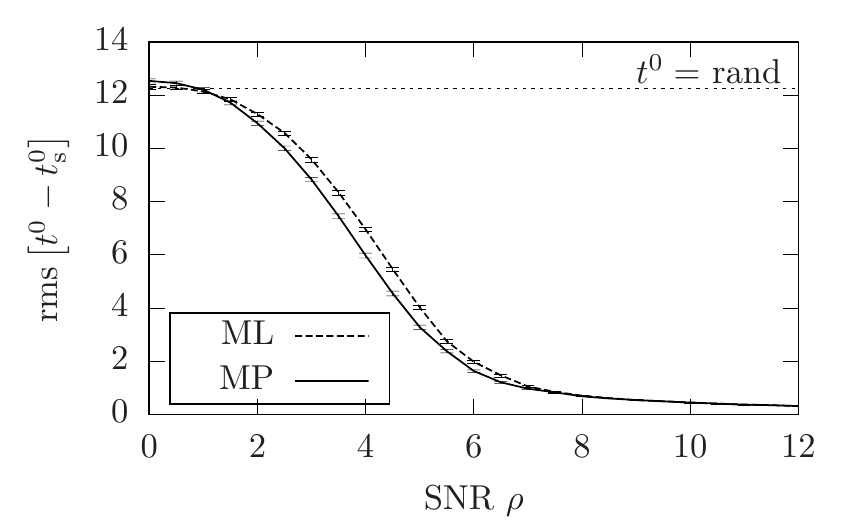}
    \includegraphics[width=0.5\textwidth]{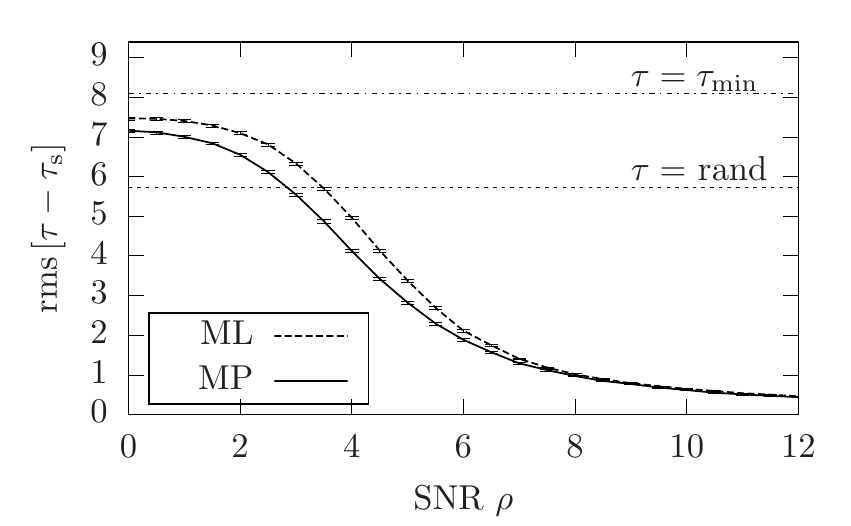}
  }
  \caption{Parameter-estimation accuracy: rms-errors on $\tStart$ (\emph{left panel})
    and $\tTau$ (\emph{right panel}) as a function of SNR for maximum-likelihood (ML) and
    maximum-posterior (MP) estimates. Each value of SNR corresponds to $N=10^4$ random
    parameter draws, and the error-bars represent $1\sigma$ jackknife error-estimates.
    The dotted horizontal line (`rand') corresponds to a uniform random guess \eqref{eq:44}, and the
    dot-dashed horizontal line corresponds to a maximally biased $\tau=\tau_{\min}$ ``guess'' \eqref{eq:97}.}
  \label{fig:paramEstimation_vs_SNR}
\end{figure*}
Similarly, parameter-estimation on the window-type $\type$ itself can be expressed as
\begin{equation}
  \label{eq:31}
  \begin{split}
    \prob{\type}{\dVx,\HypS,\Dop} &\propto \prob{\type}{\HypS}\,\int e^{\F(\dVx;\Dop,\Trans)}\,d\tStart\,d\tTau\\
     &\propto \prob{\type}{\HypS}\,\BF(\dVx;\Dop,\type)\,,
  \end{split}
\end{equation}
so the window-type specific Bayes factor \eqref{eq:25} is proportional to the relative likelihood of different
window-types $\type$.

In the frequentist framework one often uses maximum-likelihood estimators for parameter estimation,
i.e. $\{\tStart_\ML,\,\tTau_\ML\}$ such that
\begin{equation}
  \label{eq:55}
  \F(\dVx;\tStart_\ML,\tTau_\ML) = \max_{\{\tStart,\tTau\}} \F(\dVx;\tStart,\tTau)\,,
\end{equation}
for fixed Doppler-point $\Dop$ and window-type $\type$.

The following Monte-Carlo studies use rectangular transient windows within a wider range than \eqref{eq:99}, namely
\begin{equation}
  \label{eq:101}
  \rangeIII:\quad \tTau\in[0.5,14.5]\,\days\,,\;\;\tStart \in [0,30]\,\days\,.
\end{equation}
Using the transient-parameter \rangeIII, Fig.~\ref{fig:paramEstimation_pdfs} shows one example of parameter
posteriors \eqref{eq:53} and maximum-likelihood (ML) estimators \eqref{eq:55} on $\tStart$ and $\tTau$ for
injected signals with SNR $\rhoOpt=5$ and $\rhoOpt=8$, respectively.
We see in Fig.~\ref{fig:paramEstimation_pdfs} that the timescale of variations in $\prob{\tStart}{\dVx}$ is
shorter than in $\prob{\tTau}{\dVx}$. This is due to the combined effect of the wider plotted range in
$\tStart$ and the twice smaller characteristic correlation timescale in $\tStart$, as derived in Eq.~\eqref{eq:98}.

In order to study the quality of transient parameter-estimation as a function of SNR, we performed Monte-Carlo
simulations comparing maximum-posterior estimators (MP), defined as
\begin{equation}
  \label{eq:43}
  \begin{split}
    \prob{\tStart_\MP}{\dVx,\HypS} &= \max_{\tStart}\prob{\tStart}{\dVx,\HypS}\,,\\
    \prob{\tTau_\MP}{\dVx,\HypS}   &= \max_{\tTau}\prob{\tTau}{\dVx,\HypS}\,,
  \end{split}
\end{equation}
and maximum-likelihood estimators (ML) of Eq.~\eqref{eq:55}, using uniform priors on $\{\tStart,\tTau\}$ within \rangeIII,
and physical priors \eqref{eq:21} on amplitude parameters. For different fixed values of SNR we perform
$N=10^4$ simulated parameter estimates, and compute the rms errors from $\tStart - \tStart_\sig$ and $\tTau -
\tTau_\sig$,
for ML- and MP- estimators, respectively.
We also compare the results to the error of a pure random guess within the range $[t_1, t_1 + \Delta t]$, where
$\Delta t
\equiv t_2 - t_1$. For uniform priors this is
\begin{equation}
  \label{eq:44}
  (\rms{t - t_\sig})^2 = \frac{1}{\Delta t^2}\int_{t_1}^{t_2}\,d t_\sig \int_{t_1}^{t_2}\,d t\, (t -
  t_\sig)^2 = \frac{\Delta t^2}{6}\,.
\end{equation}
For the transient \rangeIII{} this yields random-guess errors of
$\rms{\tStart - \tStart_\sig}\approx 12.25\,\days$, and
$\rms{\tTau - \tTau_\sig} \approx 5.72\,\days$, which are shown in Fig.~\ref{fig:paramEstimation_vs_SNR}.
On the other hand, for a maximally biased ``guess'' of $\tTau = \tTau_{\min}$, one finds
\begin{equation}
  \label{eq:97}
  \rms{\tTau_{\min} - \tTau_\sig} = \Delta\tTau/\sqrt{3}\approx 8.08\,\days\,.
\end{equation}
The results of the parameter-estimation Monte-Carlo simulation are shown in Fig.~\ref{fig:paramEstimation_vs_SNR}.
For low SNRs of $\rho \lesssim 7$, we see that the MP-estimators
perform better than the ML estimators, while for higher SNR the estimation quality of both estimators
converges.
We note that for $\rho \rightarrow 0$ the parameter estimation on $\tStart$ converges to a random guess as
expected, but in the case of $\tTau_\ML$ we notice a substantial deviation, and to a lesser degree, also for
$\tTau_\MP$. These estimates fall closer to a maximally biased $\tTau=\tTau_{\min}$ ``guess'', which indicates
an increasing bias in the $\tTau$ estimators for low SNR, strongly favoring values close to
$\tTau_{\min}$. This surprising effect will be studied in some more detail in the following section.

\subsection{Estimation bias on timescale $\tTau$ in pure noise}
\label{sec:estimation-bias-tau}

\begin{figure}[htbp]
  \centering
  \includegraphics[width=0.5\textwidth]{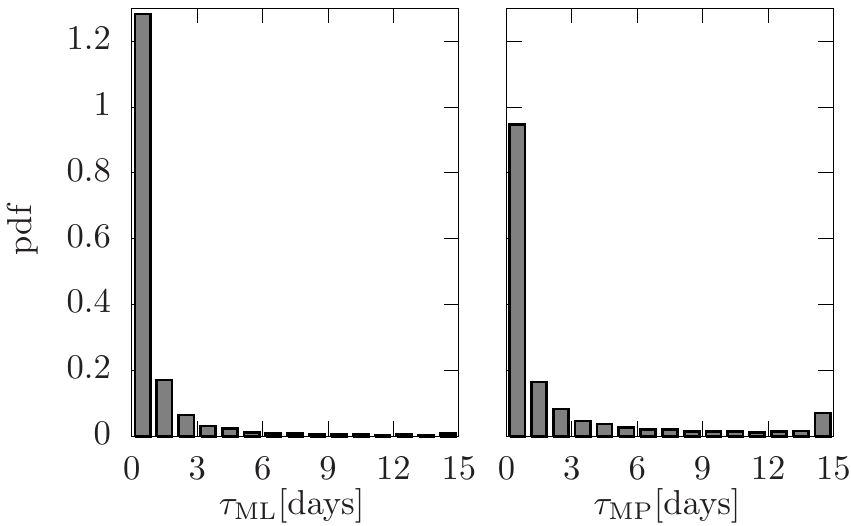}
  \caption{Parameter-estimation bias in $\tTau_\ML$ (\emph{left panel}) and $\tTau_\MP$ (\emph{right panel})
    in pure noise data (SNR=0). The plots show normalized histograms of $N=10^4$ parameter estimates of
    $\tTau$ in Gaussian random noise data.}
  \label{fig:paramEstimation_SNR0Bias}
\end{figure}
Figure~\ref{fig:paramEstimation_SNR0Bias} shows normalized histograms of the parameter estimates on $\tTau$ in
pure Gaussian noise, i.e.\ for $\rho=0$. These results confirm the estimation bias towards $\tTau_{\min}$
previously seen in Fig.~\ref{fig:paramEstimation_vs_SNR}. We have been able to trace this effect to a
surprising fundamental feature of the Gaussian random walk underlying matched filtering.
If we discretize the integration time as $T_j = j\,\Delta T$ in steps of $\Delta T$, then linearity of the
scalar product \eqref{eq:13a} implies
\begin{equation}
  \label{eq:58}
  x'_\mu(T_{j}) = x'_\mu(T_{j-1}) + \Delta x'_{\mu,j}\,,
\end{equation}
where $\Delta x'_{\mu,j}$ is independent of $x'_\mu(T_{j-1})$ and follows a Gaussian distribution with zero
mean (in the noise case $\HypN$).
We see from this expression that the amplitudes $x'_\mu$ can be interpreted as Gaussian random walks over
finite steps $\Delta T$ in the integration time. The four Gaussian random walks $\{x'_\mu\}$ are combined in the
quadratic form \eqref{eq:16} to yield $\F(T_j)$, where they are normalized by $\M^{\mu\nu}$ such that $\F$ follows a
$\chi^2$-distribution with four degrees of freedom \emph{for any} $T_j$, independently of
the random-walk step $j$.

We can therefore consider a simpler toy model, namely a 1-dimensional normalized random walk, defined as
\begin{equation}
  \label{eq:59}
  s_{n} = \frac{1}{\sqrt{n}}\, \sum_{j=1}^n \Delta x_j\,,
\end{equation}
where the $\Delta x_j$ are $n$ independent Gaussian random variables with zero mean and unit variance,
i.e. $\Delta x_j \sim \Gauss(0,1)$. The random walk \eqref{eq:59} is normalized in such a way that it follows
\emph{exactly} the same distribution at every step $n$, i.e.\ $s_n \sim \Gauss(0,1)$.
We denote as $s^2_{n_{\max}}$ the maximum of $s_n^2$ over a walk-step window $[n_1, n_2]$, i.e.
\begin{equation}
  \label{eq:60}
  s^2_{n_{\max}} \equiv \max_{n \in [n_1,n_2]} s_n^2\,,
\end{equation}
and plot the distribution of $n_{\max} \in [n_1, n_2]$ in repeated trials of such normalized random walks.
\begin{figure}[htbp]
  \centering
  \includegraphics[width=0.5\textwidth]{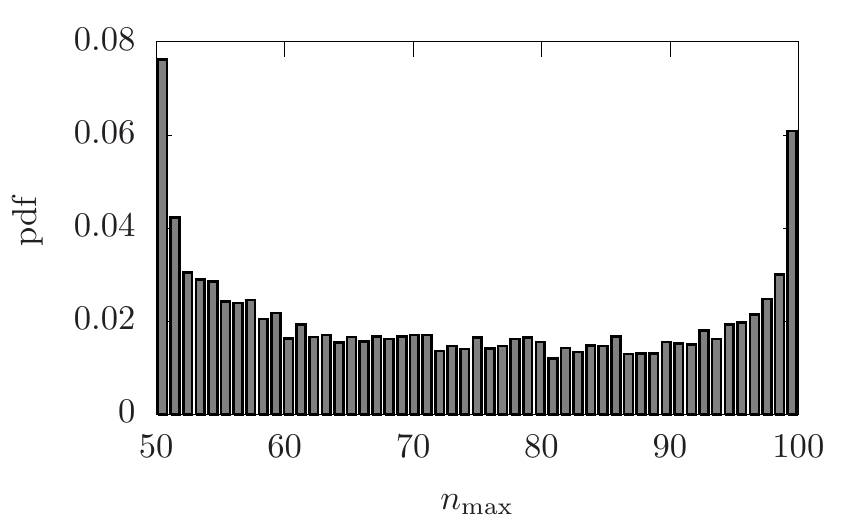}
  \caption{Distribution of random-walk index $n_{\max}$ at which the maximum of $s^2_n$ occurs ($N=10^4$ trials).}
  \label{fig:RandomWalkBias}
\end{figure}
For example, if we consider $n_1 = 50$ and $n_2 = 100$, and $N=10^4$ trials, we obtain the distribution of
maxima shown in Fig.~\ref{fig:RandomWalkBias}, which illustrates a qualitative bias towards $n_{\min}$,
and to a lesser extent $n_{\max}$, similar to what was seen in Fig.~\ref{fig:paramEstimation_SNR0Bias} for the
physical parameter estimation of $\tTau$.
This seems to be a manifestation of a well-known property of random walks, namely L\'evy's arcsine law,
which is discussed, for example, in Sec.4.2 of \cite{2010PhyA..389.4299M}.

\section{Conclusions}
\label{sec:Conclusions}

We have discussed the case for a transient-CW search, from the point of view of its
astrophysical motivation, and in order to  bridge the gap between short burst-like
signals and traditional infinite-duration CW signals.

We have introduced a simple transient signal model based on the classical CW signal, modulated by
a window function of finite support. The corresponding Bayes factor has been derived and
implemented, and we have performed Monte-Carlo simulations to compare its efficiency to the orthodox
maximum-likelihood detection method. These results show that the Bayes factor is both more sensitive and more
robust than a maximum-likelihood statistic, and it yields better parameter estimates, at similar computing cost.


The Monte-Carlo studies presented in this work have been limited to pure Gaussian noise, and it would be
important to test this method in practice on actual detector data. In particular one needs to address the
question of how to assign meaningful false-alarm probabilities to detection candidates in real detector data
(see also \cite{clarkLSC2010:_vela_glitch} for an example of these difficulties).

The search-method discussed here is currently restricted to non-repeating transient CWs, and more work would be
required to generalize this approach to allow for repeating transient-CW signals.

In addition to the fully coherent search method, we have derived the necessary formalism for a semi-coherent
transient search, which could be used to perform an all-sky, all-frequency wide parameter-space transient search,
for example running on Einstein@Home. More work is required to fully develop and implement this approach.

\section{Acknowledgments}

We thank Badri Krishnan, Curt Cutler, Christian R\"over, and Vladimir Dergachev for interesting discussions. We further thank
Holger Pletsch, Ben Owen, Nils Andersson, Ian Jones, Trevor Sidery and Eric Thrane for careful reading of the manuscript and
useful comments.

\appendix

\section{Transient search Implementation}
\label{sec:trans-search-impl}

Our numerical implementation of the Bayes factor $\BF(\dVx;\Dop)$ of Eq.~\eqref{eq:25} consists of two steps:
\begin{enumerate}
\item calculate a discretized $\F$-statistic map $\F(\dVx;\tStart,\tTau)$ over the search ranges in $\tStart$
  and $\tTau$,
\item compute $\BF(\dVx)$ by discretizing the marginalization integral in Eq.~\eqref{eq:25} as a sum.
\end{enumerate}
The following two sections provide more details about these two steps, respectively.

\subsection{Atoms-based $\F$-statistic computation}
\label{sec:atoms-based-f}

We discretize the 2-dimensional $\F$-statistic map over the search ranges
$\tStart\in [\tStart_{\min},\tStart_{\min}+\Delta\tStart]$ and $\tTau \in [\tTauMin, \tTauMin + \Delta\tTau ]$
in steps $d\tStart$ and $d\tTau$, respectively.
Namely, we compute the $\F$-matrix
\begin{equation}
  \label{eq:71}
  \F_{m\,n} \equiv \F(\dVx;\, \tStart_m,\,\tTau_n)\,,
\end{equation}

over the  $\NStart\times\NTau$ rectangular grid
\begin{equation}
  \label{eq:72}
  \begin{split}
    \tStart_m &= \tStart_{\min} + m\,d\tStart\,,\\
    \tTau_n   &= \tTau_{\min}   + n\,d\tTau\,,
  \end{split}
\end{equation}
where $\NStart = \Delta\tStart/d\tStart$ and $\NTau = \Delta\tTau / d\tTau$.

The default for these step-sizes used here are $d\tStart = d\tTau = 1800\,$s. If the shortest signals
considered are $\tTau_{\min} \sim 0.5\,\days$ long, then according to Eq.~\eqref{eq:42} the worst-case mismatch is
$\mis \lesssim d\tTau/\tTau_{\min} \approx 0.04$, i.e. a $4\%$ loss of squared SNR.

In the current implementation of the transient-CW search we use the underlying discretization of the
$\F$-statistic computation in \texttt{ComputeFStatistic\_v2} (\CFS), as described in more detail in
\cite{prix:_cfsv2}.
Namely the $\F$-statistic \eqref{eq:16} is computed from $\M'_{\mu\nu}$ and $x'_\mu$ of
Eqs.~\eqref{eq:13a},\eqref{eq:13b}, which are approximated as sums
\begin{equation}
  \label{eq:73}
  \begin{split}
    x_\mu      &\approx \sum_{i=1}^{\NSFT} \win_i \, x_{\mu,i}\,,\\
    \M_{\mu\nu} &\approx \sum_{i=1}^{\NSFT} \win_i^2 \, \M_{\mu\nu,i}\,,
  \end{split}
\end{equation}
in terms of the $\F$-statistic ``atoms''
\begin{equation}
  \label{eq:74}
  \begin{split}
    x_{\mu,i}    &\equiv 2\sum_X S^{-1}_X \int_{t_i}^{t_i + \TSFT}  x^X(t)\, h^X_\mu(t)\,dt\,,\\
    \M_{\mu\nu,i} &\equiv 2\sum_X S^{-1}_X \int_{t_i}^{t_i + \TSFT} h^X_\mu(t)\,h^X_\nu(t)\,dt\,,
  \end{split}
\end{equation}
where $\TSFT$ is the length of the short Fourier transforms (SFTs) that are used as input data,
typically $\TSFT=1800\,$s. In the above expressions we implicitly assumed that the transient-window function
$\win(t)$ varies slowly and can be approximated as constant over the timescale $\TSFT$.
For any chosen Doppler position $\Dop$, the code first computes the $\NSFT$ atoms
$\{x_{\mu,i},\,\M_{\mu\nu,i}\}_{i=1}^{\NSFT}$ over the whole observation time of interest,
which are also the primary input to this implementation of the standard CW $\F$-statistic.
The $\F$-statistic value $\F_{m\,n}$ for any particular transient parameters $\{\tStart_m,\,\tTau_n\}$
is then computed from the corresponding partial sums in Eq.~\eqref{eq:73}.
This approach allows for an efficient computation of the $\F_{m\,n}$ map in the
case of a rectangular window function $\win_\rect$ of Eq.~\eqref{eq:2}: going from $\tTau_n$ to $\tTau_{n+1}$
can be achieved by a single extra addition, namely
\begin{equation}
  \label{eq:75}
  x_{\mu}(\tStart_m,\tTau_n) = x_{\mu}(\tStart_m,\tTau_{n-1}) + x_{\mu,i_{1}}\,,
\end{equation}
(and similarly for $\M_{\mu\nu}$), where $i_1$ is the atom-index corresponding to the time-step  $\tStart_m +
\tTau_{n}$ (assuming for simplicity that $d\tTau=\TSFT$).

In the case of the exponential transient window $\win_\ex$ of Eq.~\eqref{eq:3}, the whole sum in Eq.~\eqref{eq:73} needs
to be recomputed for every matrix-element $m,n$, as the window-function provides different weights at every point.

\subsection{Transient marginalization integrals}
\label{sec:trans-marg-integr}

Given the transient matrix $\F_{m\,n}$, we can now turn the integrals in Eqs.~\eqref{eq:25} and \eqref{eq:53} into
simple sums.
A minor subtlety arises because of the potential numerical problems of expressions like $e^\F$, which overflow
in double precision for values $\F>709$. Such values are easily possible for noisy non-Gaussian data with
line-artifacts or for strong injected signals.
It is therefore numerically safer to rewrite these sums using the discretized maximum-likelihood statistic
value \eqref{eq:27}, namely
\begin{equation}
  \label{eq:77}
  \Fmax \approx \max_{m\,n} \F_{m\,n}\,,
\end{equation}
and write $\BF$ of Eq.~\eqref{eq:25} as a discretized sum in the form
\begin{equation}
  \label{eq:78}
  \log\BF(\dVx;\,\Dop) \approx \Fmax + c_0 + \log \sum_{m=1}^{\NStart} \sum_{n=1}^{\NTau} e^{\Delta\F_{m\,n}} \,,
\end{equation}
where we defined $c_0 \equiv \log(70/(\rhohMax^4\, \NStart\,\NTau))$, and
\begin{equation}
  \label{eq:79}
  \Delta\F_{m n} \equiv \F_{m n} - \Fmax \le 0\,.
\end{equation}
For large values of $\Fmax$, some terms $e^{\Delta\F_{m n}}$ can now \emph{underflow} to zero, but this poses
no numerical problems because these contributions were negligible anyway.
Similarly, we can compute the parameter posteriors \eqref{eq:53} as
\begin{align}
  \label{eq:76}
  & \prob{\tStart_m}{\dVx,\HypS} \propto \sum_{n=1}^{\NTau} e^{\Delta\F_{m\,n}}\,,\\
  & \prob{\tTau_n}{\dVx,\HypS} \propto \sum_{m=1}^{\NStart} e^{\Delta\F_{m\,n}}\,,
\end{align}
where $\Fmax$ only affects the normalization and has been dropped.

\subsection{Computing cost}
\label{sec:CompCost}

In order to be able to plan which types of searches can be performed with a reasonable investment of computing
cost, it is useful to have a rough computing-cost model that allows one to predict the expected run time of any search.

Let us start with the underlying $\F$-statistic implementation \CFS{} which is used in our current coherent
transient-search implementation described in Sec.~\ref{sec:atoms-based-f}. Note that more efficient $\F$-statistic
algorithms do exist, based on resampling and FFT techniques
\cite{2002PhRvD..65d2003A,2010PhRvD..81h4032P}. However, the ``Williams-Schutz'' method
\cite{williams99:_effic_match_filter_algor_detec} currently implemented in \CFS{} is well suited
to our purpose, as it is already based on computing the atoms $\{x_{\mu,i},\M_{\mu\nu,i}\}$ over
SFTs. Nevertheless it will be interesting to study possibly more efficient transient-CW implementations based
on resampling and FFT techniques.

We can break up the total computing time $\cost_{B}^{(1)}$ for $\BF(\dVx;\Dop)$ at one Doppler point $\Dop$ as follows:
\begin{enumerate}
\item the time $\cost_\atoms$ to compute the atoms \eqref{eq:74},
\item the time $\cost^\type_\Fmap$ to compute the $\F$-map \eqref{eq:71} over $\{\tStart,\tTau\}$ for given window-type $\type$, and
\item the time $\cost_\marg$ to marginalize over the $\F$-map to obtain the Bayes factor $\BF(\dVx;\Dop)$,
  using Eq.~\eqref{eq:78}
\end{enumerate}

\textbf{Atoms cost:} the time $\cost_\atoms$ to compute for one Doppler position $\Dop$
the $\NSFT = \Tdata /
\TSFT$ atoms using \CFS{} is simply
\begin{equation}
  \label{eq:80}
  \cost_\atoms = \cost_0\,\, \frac{\Tdata}{\TSFT} \,,
\end{equation}
where $\cost_0$ is a machine-dependent timing constant.

$\mbf{\F}$-\textbf{map cost:} the time $\cost^\type_\Fmap$  to compute the $\NStart\times\NTau$ matrix of values
$\{\F_{m\,n}\}$, where $\NStart = \Delta\tStart/d\tStart$, and $\NTau = \Delta\tTau/d\tTau$. This time can be
expressed as
\begin{equation}
  \label{eq:81}
  \cost^\type_\Fmap = \sum_{m=1}^{\NStart}\sum_{n=1}^{\NTau} \cost^\type_{m\,n}\,,
\end{equation}
where $\cost^\type_{m\,n}$ is the time required to compute $\F_{m\,n}$ for particular values $m,n$, which
depends crucially on the type $\type$ of window-function.

Let us first consider the exponential transient-window \eqref{eq:3}, which corresponds to the generic case where
no special optimizations can be used.
Note that at the smallest timescale index, $n=1$, we already had to sum all atoms corresponding to a timescale
$\tTau_{\min}$. However, for every matrix element we need to re-compute this sum, as the window-weights will
be different every time.
This introduces a constant computing-time offset $\cost_{n=0} = \cost_\ex \,\tTau_{\min}$, where $\cost_\ex$ is the machine-dependent time to
compute all atom-additions and window-weights within one unit-time. From this we can express the cost for
computing $\F_{m\,n}$ as $\cost^\ex_{m\,n} = \cost_\ex \, ( \tTau_{\min} + n\, d\tTau )$,
which does not depend on the start-time index $m$. Therefore we obtain the total $\F$-map cost as
\begin{equation}
  \label{eq:84}
  \cost^\ex_\Fmap \approx \cost_\ex \, \frac{\Delta\tStart}{d\tStart}\,\frac{\Delta\tTau}{d\tTau} \, \frac{(\tTau_{\min} + \Delta\tTau/2)}{\TSFT} \,,
\end{equation}
which is quadratic in $\Delta\tTau$ (and where we assumed \mbox{$\NTau\gg1$}).

In the case of a rectangular transient-window, various sums are closely related and we can use the
optimization of Eq.~\eqref{eq:75} to reduce the computing cost. Namely, every step $n\rightarrow n+1$ in the
timescale $\tTau$ adds just the cost of one extra time-step $d\tTau$. This cost is $\cost_\rect\,d\tTau$,
where $\cost_\rect$ is the machine-dependent time to do all sums for a unit timescale in the case of a
rectangular window.
For every start-time index $m$, we need to compute the sums up to $\tau_{\min}$ first, costing
$\cost_\rect\,\tau_{\min}$. Summarizing, we can express the accumulated computing cost
$\cost^\rect_m =
\sum_n \cost^\rect_{m\,n}$ \emph{per line} of the $\F_{m\,n}$ matrix as
$\cost^\rect_m = \cost_\rect \, \tTau_{\min} + \cost_\rect\,\NTau d\tTau = \cost_\rect\,\tau_{\max}$.
The total $\F$-map cost is therefore
\begin{equation}
  \label{eq:85}
  \cost^\rect_\Fmap = \cost_\rect \, \frac{\Delta\tStart}{d\tStart} \, \frac{(\tTau_{\min} + \Delta\tTau)}{\TSFT}\,.
\end{equation}

\textbf{Bayes-factor marginalization:} the marginalization \eqref{eq:78} is a simple sum over the
exponentiated $\F$-map matrix, and so we can directly write the marginalization cost as
\begin{equation}
  \label{eq:86}
  \cost_\marg = \cost_1\,\frac{\Delta\tStart}{d\tStart}\,\frac{\Delta\tTau}{d\tTau}\,,
\end{equation}
where $\cost_1$ is the machine-dependent cost of exponentiation and summation of real numbers. Note that
exponentiation is a very costly operation, and so we use a lookup-table approximation to reduce
this cost.

The total computing cost $\cost_{B}^{(1)}$ for the transient Bayes factor at one Doppler point $\Dop$ is
now expressible as
\begin{equation}
  \label{eq:87}
  \cost_{B}^{(1)} = \cost_\atoms + \cost^\type_\Fmap + \cost_\marg\,,
\end{equation}
and for a search over $N_\Dop$ Doppler points, this would simply extend as
\begin{equation}
  \label{eq:88}
  \cost^{(N_\Dop)}_{B} = N_\Dop\, \cost^{(1)}_{B}\,,
\end{equation}
where the extra cost of summing these Bayes-factors will be negligible.

We have verified that these timing models are a good description of the actual performance of the code
\footnote{Note that the atoms-cost $\cost_0$ per SFT reported here refers to a standard (-O2) build,
  while highly optimized versions of this code running on Einstein@Home can achieve more than 10x faster
  performance.}
by varying the parameters $\NSFT$, $\Delta\tStart$ and $\Delta\tTau$, and by fitting the measured times to the
model.
This yields the following timing constants on an Intel Core2 Duo CPU with 2.60~GHz (Lenovo T61p):
\begin{equation}
  \label{eq:17}
  \begin{split}
    \cost_0    &= 1.4\times 10^{-6}\;\secs,  \quad   \cost_1   = 2.8\times10^{-8}\;\secs\,.\\
    \cost_\rect &= 4.2\times10^{-8}\;\secs, \quad     \cost_\ex  = 1.3\times10^{-7}\;\secs\\
\end{split}
\end{equation}

If we target a known pulsar with a transient-CW search using data from 2 detectors spanning one year ($\NSFT=35,000$), with a timescale
range of $\tTau \in [0.5,\,14.5]\,\days$, using step-sizes
$d\tStart = d\tTau = \TSFT = 1800\,\secs$, we obtain an estimated computing costs of
\begin{equation}
  \label{eq:18}
  \begin{split}
    \cost_\atoms     &\approx 0.05\,\secs, \quad  \cost_\marg \approx 0.3\,\secs\\
    \cost_\Fmap^\rect &\approx 0.5\,\secs,  \quad  \cost_\Fmap^\ex \approx 540\,\secs\,.
  \end{split}
\end{equation}
We see that such a targeted search would be easy to perform for all interesting pulsars, even using the much slower
transient exponential window. Comparing this transient-CW search to a coherent CW search over one
year, we notice that the cost per template $\Dop$ is about 18 times higher than the CW search for a
rectangular window, and about $10^4$ times higher for the exponential window.
Wide parameter-space transient-CW searches will therefore be severely limited by computing resources, and a
semi-coherent transient search method as discussed in Sec.~\ref{sec:semi-coher-trans} will be required.

\subsection{Synthesizing Monte-Carlo draws}
\label{sec:synth-monte-carlo}

Following the method used in \cite{2009CQGra..26t4013P}, we have implemented an efficient Monte-Carlo
simulation method, by avoiding the generation of the primary data-input of the search code (time-series or short
Fourier transforms), and instead synthesizing higher-level secondary data-input to the transient-search
functions directly. In our case, we synthesize the $\NSFT$ \emph{atoms}
$\{x_{\mu,i},\,\M_{\mu\nu,i}\}$, which are the intermediate input-data to the transient-search function.
This approach is very economical in computing resources and allows us to generate large numbers of Monte-Carlo
draws in a very short time on a single machine.
We draw signal parameters $\{\Amp_\sig, \Dop_\sig, \Trans_\sig\}$ according to the priors, and from this we can
compute the deterministic signal-atoms $s_{\mu,i}$ of Eq.~\eqref{eq:34} and the antenna-pattern
atoms $\M_{\mu\nu,i}$ of Eq.~\eqref{eq:74}.
The noise-atoms $n_{\mu,i}$ are Gaussian random variables with zero mean and covariance
matrix $\M_{\mu\nu,i}$. These can be generated from uncorrelated Gaussian variates, e.g.\ by using
a Cholesky decomposition on $\M_{\mu\nu,i}$. The data-atoms are then simply
$x_{\mu,i} = n_{\mu,i} + s_{\mu,i}$, according to Eq.~\eqref{eq:28}. Note that we have assumed the
transient-window function to be constant on the atoms timescale $\TSFT$, and therefore we can synthesize
standard non-transient CW atoms. The transient-window function is applied when computing
the $\F$-statistic map, cf.\ Sec.~\ref{sec:atoms-based-f}.
All Monte-Carlo simulations used the ``Mersenne Twister'' random-number generator \texttt{gsl\_rng\_mt19937}
from GSL\cite{libgsl}.

\bibliography{biblio}

\end{document}